# Exceptional piezoelectricity, high thermal conductivity and stiffness and promising photocatalysis in two-dimensional MoSi$_2$N$_4$ family confirmed by first-principles

Bohayra Mortazavi*[a,b], Brahmanandam Javvaji[a,#], Fazel Shojaei[c,#], Timon Rabczuk[d], Alexander V. Shapeev[e] and Xiaoying Zhuang**[a,b,d]

[a]*Chair of Computational Science and Simulation Technology, Institute of Photonics, Department of Mathematics and Physics, Leibniz Universität Hannover, Appelstraße 11,30157 Hannover, Germany.*
[b]*Cluster of Excellence PhoenixD (Photonics, Optics, and Engineering–Innovation Across Disciplines), Gottfried Wilhelm Leibniz Universität Hannover, Hannover, Germany.*
[c]*Department of chemistry, Faculty of sciences, Persian Gulf University, Bushehr 75169, Iran.*
[d]*College of Civil Engineering, Department of Geotechnical Engineering, Tongji University, 1239 Siping Road Shanghai, China.*
[e]*Skolkovo Institute of Science and Technology, Skolkovo Innovation Center, Nobel St. 3, Moscow 143026, Russia.*

## Abstract

Chemical vapor deposition has been most recently employed to fabricate centimeter-scale high-quality single-layer MoSi$_2$N$_4$ (*Science; 2020;369; 670*). Motivated by this exciting experimental advance, herein we conduct extensive first-principles based simulations to explore the stability, mechanical properties, lattice thermal conductivity, piezoelectric and flexoelectric response, and photocatalytic and electronic features of *MA$_2$Z$_4$* (M= Cr, Mo, W; A= Si, Ge; Z=N, P) monolayers. The considered nanosheets are found to exhibit dynamical stability and remarkably high mechanical properties. Moreover, they show diverse electronic properties from antiferromagnetic metal to half metal and to semiconductors with band gaps ranging from 0.31 to 2.57 eV. Among the studied nanosheets, the MoSi$_2$N$_4$ and WSi$_2$N$_4$ monolayers yield appropriate band edge positions, high electron and hole mobilities, and strong visible light absorption, highly promising for applications in optoelectronics and photocatalytic water splitting. The MoSi$_2$N$_4$ and WSi$_2$N$_4$ monolayers are also predicted to show outstandingly high lattice thermal conductivity of 440 and 500 W/mK, respectively. For the first time we show that machine learning interatomic potentials trained over small supercells can be employed to examine the flexoelectric and piezoelectric properties of complex structures. As the most exciting finding, WSi$_2$N$_4$, CrSi$_2$N$_4$ and MoSi$_2$N$_4$ are found to exhibit the highest piezoelectric coefficients, outperforming all other-known 2D materials. Our results highlight that *MA$_2$Z$_4$* nanosheets not only undoubtedly outperform the transition metal dichalcogenides group but also can compete with graphene in many applications in nanoelectronics, optoelectronic, energy storage/conversion and thermal management systems.

Corresponding authors: *bohayra.mortazavi@gmail.com; **zhuang@iop.uni-hannover.de
[#]These authors contributed equally

## 1. Introduction

Graphene's first experimental realization in 2004 [1–3] and confirmation of its ultrahigh thermal conductivity [4,5], mechanical strength [6] and carrier mobility [7] along with outstanding electronic and optical properties [8–11], initiated extending and continuous interests toward two-dimensional (2D) materials. The pristine graphene with full-sp$^2$ carbon atoms does not however exhibit an electronic band gap. The semimetal electronic nature of defect-free graphene limits its application in nanoelectronics and nano-optics. In fact, having a proper electronic band gap is a critical requirement for the majority of rapidly growing technologies. During the last decade numerous techniques have been proposed for the band gap opening in graphene, like the formation of point defects [12–16], patterned cuts [17,18], stretching [19–23], and doping [24–28]. Nonetheless all the aforementioned techniques require additional processing after the fabrication of graphene and thus result in increased complexity and production cost as well. This way, for practical applications it is more recommended to fabricate an intrinsic 2D semiconductor rather than to engineer the graphene's electronic structure. This issue motivated the design and fabrication of novel 2D semiconductors, such as the transition metal dichalcogenides [29,30], indium selenide [31] and phosphorene [32,33]. In comparison with graphene, the majority of 2D semiconductors exhibit distinctly lower mechanical strength and thermal conductivity. In fact, one of the most appealing features of graphene is its superior thermal conductivity that outperforms all other materials and propose it as unique candidate for the thermal management systems [34–37]. This way, a 2D semiconductor with high thermal conductivity is greatly appealing to tackle the common overheating concern in nanoelectronics. Moreover, the application of novel nanomaterials in energy storage/conversion systems is another very active and attractive field of research [38–41]. In this regard, piezoelectricity and flexoelectricity are currently playing critical roles in many advanced technologies [42–46]. Therefore the design of novel 2D materials with appealing piezoelectricity and flexoelectricity are highly appealing to expand the practical application of 2D materials.

In line of continuous expansion of 2D semiconductors, in a latest study by Hong and coworkers [47], they succeeded in the first experimental realization of large-area MoSi$_2$N$_4$ monolayer by incorporating silicon during chemical vapor deposition growth of molybdenum nitride. This novel 2D system was found to be a semiconductor with remarkable tensile strength and high carrier mobility as well [47]. As an exciting fact, this latest accomplishment paves the path for the experimental realization of an extensive family of *MA$_2$Z$_4$* nanosheets, in which M is an early transition metal (Mo, W, V, Nb, Ta, Ti, Zr, Hf or Cr), A is either Si or Ge and Z can be N, P or As [48]. Their unique sandwich structure with the possibility of altering the different layers' composition create vast opportunities to reach diverse properties [48].

Motivated by this latest experimental accomplishment by Hong *et al.* [47], in this study our objective is to examine the stability, mechanical response, lattice thermal conductivity, piezoelectricity and flexoelectricity and electronic features of twelve different *MA$_2$Z$_4$* (M= Cr, Mo, W; A= Si, Ge; Z=N, P) monolayers. For every composition we consider two different structures and evaluate their energetic and dynamical stability. The considered representative lattices in this work can establish a comprehensive vision on the intrinsic properties of $MA_2Z_4$ nanosheets. We explored the intrinsic properties by conducting density functional theory (DFT) based calculations. For the most stable configurations the acquired results reveal semiconducting electronic nature with band gaps ranging from 0.8 to 2.6 eV. Our first-principles results confirm remarkably good mechanical properties and thermal conductivity of this novel class of 2D materials. In particular, the $MoSi_2N_4$ and $WSi_2N_4$ monolayers are predicted to show outstandingly high lattice thermal conductivity of 440 and 500 W/mK, and elastic modulus of 487 and 506 GPa, respectively. These aforementioned monolayers are also found to exhibit appropriate band edge positions, high electron and hole mobilities, and strong visible light absorption, highly promising for the applications in optoelectronics and photocatalytic water splitting. As the first study, we show that machine learning interatomic potentials can be effectively employed to evaluate piezoelectric and flexoelectric responses of 2D systems. Notably, we predict that the $WSi_2N_4$, $CrSi_2N_4$ and $MoSi_2N_4$ monolayers exhibit the highest piezoelectric coefficients, outperforming all other-known 2D materials. The presented first-principles results provide a comprehensive vision on the critical properties of $MA_2Z_4$ 2D materials family and highlight their very attractive properties for the design of novel nanoelectronics, optoelectronics, thermal management and energy conversion/storage systems.

## 2. Computational methods

Density functional theory (DFT) simulations in this work are conducted using the *Vienna Ab-initio Simulation Package* (VASP)[49–51] with generalized gradient approximation (GGA) and Perdew–Burke–Ernzerhof (PBE)[52] functional considering an energy cutoff of 500 eV for the plane waves. Energy minimization is achieved with the conjugate gradient approach with the convergence criteria of $10^{-6}$ eV and 0.002 eV/Å for the energy and forces, respectively, employing a 14×14×1 Monkhorst-Pack[53] k-point grid. Spin polarized calculations are conducted to examine the possibility of magnetism in these structures. Mechanical properties are examined by conducting uniaxial tensile simulations. Density functional perturbation theory (DFPT) simulations over 4×4×1 supercells are carried out to acquire the force constants. Phonon dispersion relations are then acquired employing the PHONOPY code [54] with the DFPT results as inputs. For the analysis of electronic and optical features we consider the convergence criterion of $10^{-5}$ eV for the electronic self-consistent-loop and use a finer k-point grid of 12×12×1. Since PBE/GGA underestimates the position of conduction band maximums and systematically underestimate the band gap, the screened hybrid functional of HSE06 [55] is employed to provide more accurate estimations for the electronic and optical properties.

Charge carrier mobilities are calculated from the deformation potential approximation [56] via: $e\hbar^3 C_{2D}/KT m_e^* m_d (E_l^i)^2$, in which $\hbar$ is the reduced Planck constant, $K$ is the Boltzmann constant, $C_{2D}$ and $m^*$ are the elastic modulus and the effective mass of the carrier along the transport direction, respectively, $m_d$ is the average effective mass along both planar directions and $E^i_l$ mimics the deformation energy constant of the carrier due to phonons for the i-th edge band along the transport direction. We examine the light absorption properties of the two systems by calculating their frequency-dependent dielectric matrix, neglecting the local field effects. The imaginary part ($\varepsilon_2$) of the frequency-dependent dielectric matrix is using the following equation [57]:

$$\varepsilon_{\alpha\beta}^{2}(\omega) = \frac{4\pi^2 e^2}{\Omega} \lim_{q\to 0} \frac{1}{q^2} \sum_{c,v,\mathbf{k}} 2w_k \delta(\varepsilon_{c\mathbf{k}} - \varepsilon_{v\mathbf{k}} - \omega) <u_{c\mathbf{k}+\mathbf{e}_\alpha q}|u_{v\mathbf{k}}><u_{c\mathbf{k}+\mathbf{e}_\beta q}|u_{v\mathbf{k}}>^*$$ (1)

where indices $c$ and $v$ refer to conduction and valence band states, respectively; $w_k$ is the weight of the $k$-point; and $u_{ck}$ is the cell periodic part of the orbitals at the $k$-point. The real part ($\varepsilon_1$) of the tensor is obtained from the Kramers-Kronig relation [57]. The absorption coefficient is calculated from the following:

$$\alpha(\omega) = \sqrt{2}\omega \left[\frac{\sqrt{\varepsilon_1^2+\varepsilon_2^2}-\varepsilon_1}{2}\right]^{1/2}$$ (2)

In this work, machine-learning interatomic potentials are developed to evaluate the phononic properties, lattice thermal conductivity, piezoelectric and flexoelectric responses of $MA_2Z_4$ monolayers. As the first study, in this work we extend the application of machine-learning interatomic potentials for the modeling of piezoelectricity and flexoelectricity. To this aim we train moment tensor potentials (MTPs)[58], which have been proven as accurate and computationally efficient models for describing the atomic interactions [59–61]. The training sets are prepared by conducting ab-initio molecular dynamics (AIMD) simulations over 4×3×1 supercells with 2×2×1 k-point grids. AIMD simulations are carried out at 50 and 600 K, each for 1000 time steps. Half of the AIMD trajectories are selected to create the training sets. MTPs are then trained passively with the same procedure explained in our earlier studies [62,63] using the MLIP package [64]. The PHONOPY code [54] is employed to obtain phonon dispersion relations and harmonic force constants over 5×5×1 supercells using the trained MTPs for the force calculations [62,63]. Anharmonic interatomic force constants are calculated using the trained MTPs over 5×5×1 supercells with taking into account the interactions with eights nearest neighbors. Lattice thermal conductivity is acquired by conducting the full iterative solutions of the Boltzmann transport equation using the ShengBTE [65] package, with harmonic and anharmonic interatomic force constants as inputs. Isotope scattering is considered in the Boltzmann transport solution in order to estimate the thermal conductivity of naturally occurring structures. For every structure the convergence with respect to the $q$-grid is tested and depending on the structure between 61×61×1 to 101×101×1 $q$-grids are employed. Complete computational details for the MTP/ShengBTE coupling can be found in our earlier study [63].

In order to calculate the electrical polarization due to mechanical deformation, we use a combination of short-range bonding interactions with long-range charge-dipole (CD) interactions. In this work trained MTPs are used to describe short-range interactions. In the CD model [66,67], each atom is assumed to carry a charge $q$ and dipole moment $p$. The estimation of $q$ and $p$ variables for each atom requires the parameter $R$ (related to polarizability) and $\chi$ (electron affinity). $\chi$ values for atom types N, Cr, P, Mo, W, Ge and Si are -0.07 [68], 0.676 [69],

0.746 [70], 0.747 [69], 0.816 [71], 1.233 [72] and 1.389 [73], respectively. To quantify $R$ value for each atom type in MA$_2$Z$_4$, we perform DFT simulations for different sized atomic systems and the isotropic polarizability values from DFT ($\alpha_{\text{DFT}}$) are evaluated. With these atomic configurations and assuming $R$ varies between $0.1$ to $1.6$ Å for each atom type, the total polarizability ($\alpha_{\text{CAL}}$) from the CD model can be estimated. After that, establishing a close match between $\alpha_{\text{DFT}}$ and $\alpha_{\text{CAL}}$ results the parameter $R$ for each atom type in MA$_2$Z$_4$ monolayer. The GAUSSIAN software [74] is utilized to perform the DFT simulations for measuring $\alpha_{\text{CAL}}$. The calculation procedure of CD parameters and the implementation of the CD model incorporating into the atomic configuration under mechanical deformation are discussed in the earlier works in details [75,76].

We evaluate the in-plane piezoelectric coefficients and out-of-plane bending coefficients for a square sheet of MA$_2$Z$_4$ monolayers of dimensions of nearly 80×80 Å$^2$. *Large-scale Atomic/Molecular Massively Parallel Simulator* (LAMMPS) [77] package is employed. To determine the in-plane piezoelectric coefficient, the following displacement field of $u_y = K_t\, y$ is used to the atomic system, where $y$ represents the atomic coordinate in the $y$ direction and $K_t$ represents the given in-plane strain. For the bending deformation, we apply displacement field ($u_z$) of $u_z = \frac{1}{2} K_b y^2$ where $\frac{1}{2} K_b$ represents the given strain gradient to the bending plane. Fig. 1 illustrates the schematics of loading schemes for both tensile and bending deformations. The axes in Fig. 1 show that sheet's center is not having a displacement. The blue to red color gradient in the upper right inset represents the application of displacement $u_y$ according to the displacement field. The symmetrical color gradient from center to edges along the $y$ direction indicates the applied displacement $u_z$. Here the diamond and hexagonal markers indicate the variation of strain components $\epsilon_{yy}$ (tensile) and $\epsilon_{zy}$ (bending). During deformation, the numerical values of these components for each atom are calculated according to previous references [78,79]. Once the deformation is prescribed, the boundary atoms are kept fixed whereas the interior atoms are allowed to relax using the constant temperature simulations (NVT ensemble). After that, the point charges and dipole moments are found for each atom by the CD model. The total polarization of the atomic system is calculated $\mathbf{P} = \frac{\sum_{i=1}^{n} \mathbf{p}_i}{A}$, where $A$ is the area and $n$ is the total number of atoms in the interior (excluding the atoms near the edges) atomic system. This polarization has contributions from both piezoelectric and flexoelectric effects, which is

$$P_\alpha = d_{\alpha\beta\gamma}\epsilon_{\beta\gamma} + \mu_{\alpha\beta\gamma\delta}\frac{\partial \epsilon_{\beta\gamma}}{\partial r_\delta}, \qquad (3)$$

where $d_{\alpha\beta\gamma}$ and $\mu_{\alpha\beta\gamma\delta}$ are the piezoelectric and flexoelectric coefficients, respectively. Term $\frac{\partial \epsilon_{\beta\gamma}}{\partial r_\delta}$ represents the strain gradient to the atomic system. From the variation between polarization to strain and polarization to strain gradient, we estimate the piezo and flexoelectric coefficients, respectively.

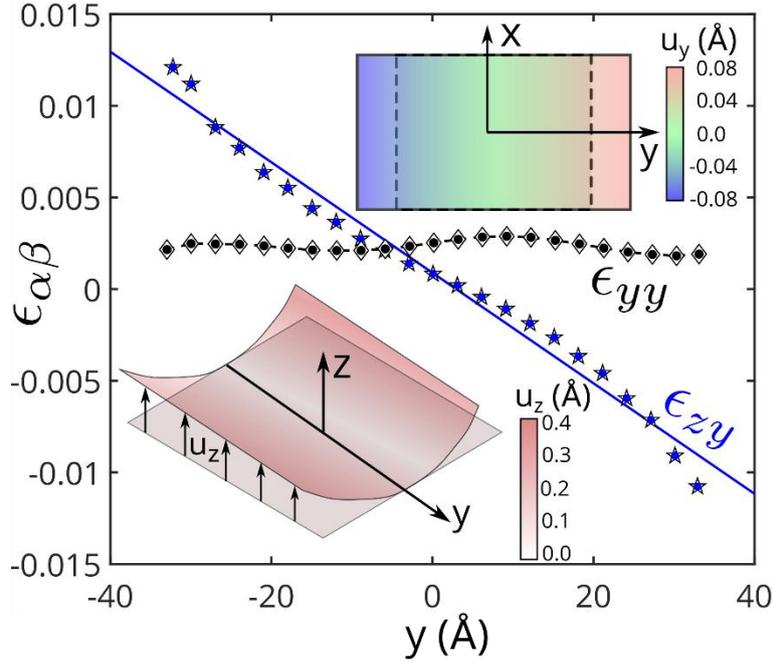

Fig. 1, The distribution of strain along y direction for the MoSi$_2$P$_4$ sheet in both tensile (Kt = 0.002) and bending (Kb = 0.0005 Å$^{-1}$) deformations. The upper right and lower left corner insets depict the schematic for the tensile and loading schemes, respectively. The color coding indicates the range for the displacements uy and uz associated with the deformation.

## 3. Results and discussions

We first introduce the structural features of *MA$_2$Z$_4$* (M= Cr, Mo, W; A= Si, Ge; Z=N, P) monolayers investigated in this work. In accordance with TMD structures, two different lattices are considered for each composition, namely, 2H and 1T phases, in which M atom has trigonal prismatic and octahedral coordination, respectively. As illustrated in Fig. 2, each monolayer can be viewed as a 2H/1T-MoS$_2$-like MZ$_2$ (MoN$_2$, taking MoSi$_2$N$_4$ as a demonstrator) monolayer sandwiched in-between two buckled honeycomb AZ (SiN) layers. The three layers are stacked on each other in a way that for 2H phase, M atom is located at the center of a trigonal prism building block with six A atoms, and the MZ$_2$ (MoN$_2$) layer is bonded to AZ (SiN) layers via vertically aligned A-Z (Si-N) bonds. The same holds for 1T phase with a difference that M atom can be thought of at the center of center of by 60° twisted trigonal prism building block with six A atoms. 2H- and 1T-*MA$_2$Z$_4$* compounds have a hexagonal primitive unitcell with space groups of P-6m2 (No. 187) and P3m1 (No. 156), respectively. Table 1 lists the structural, energetic, electronic, and magnetic properties of 2H- and 1T-*MA$_2$Z$_4$* compounds. The geometry optimized structures and their corresponding PAW potentials in VASP native format are included in the data availability section. We first examine the dynamical stability by considering the calculated phonon dispersion relations, as illustrated in Fig. 3 and Fig. S1 for 2H and 1T phases, respectively. It is clear that for all considered 2H-*MA$_2$Z$_4$* monolayers the phonon dispersion relations are free of any imaginary frequency, confirming the dynamical stability. On the other side as illustrated in Fig. S1, except for the case of 1T-CrGe$_2$N$_4$, for all other considered 1T *MA$_2$Z$_4$* lattices two of the acoustic modes show imaginary frequencies, questioning their dynamical stability. Before we discuss the relative energetic stabilities of 2H and 1T phases of each of $MA_2Z_4$ compounds, their possible magnetic ground is carefully examined. To study possible magnetism in $MA_2Z_4$ monolayers, we investigate three competing states, namely, nonmagnetic (NM), ferromagnetic (FM), and antiferromagnetic (AF). Because there is only one M atom in the unit cell of $MA_2Z_4$ monolayers, we constructed a rectangular cell with two M atoms for each one to accommodate the AF spin order. The calculated total energies show that all $MoA_2Z_4$ and $WA_2Z_4$ monolayers possess a NM ground state. $CrA_2Z_4$ monolayers however exhibit diverse magnetic characteristics. According to our results summarized in Table 1, 2H-$CrSi_2N_4$, 2H-$CrSi_2P_4$, and 2H-$CrGe_2N_4$ are found to have a NM ground state, 1T-Cr*Ge₂N₄* favors a FM spin order and it carries a net moment of 2$\mu_B$/cell, however, for the four rest monolayers of 1T-Cr*Si₂N₄*, 1T-Cr*Si₂P₄*, and 1T-Cr*Ge₂P₄* and 2H-CrGe₂P₄ AF spin order with net 0$\mu_B$/cell is favored. We found that in all cases with FM and AF spin order, local magnetic moment is almost entirely located on the Cr. Previous theoretical studies have indicated that 2H-phase of single layer pristine $MX_2$, where M= Cr, Mo, W and X= O, S, Se, Te is energetically more favorable than other configurations [80,81]. An almost similar stability trend is observed for 2H- and 1T-$MA_2Z_4$, with one exception in which FM 1T-CrGe₂N₄ is more stable than its corresponding NM 2H phase by 0.007 eV/atom. Small relative energies ($E_{rel}$) (0.032 eV/atom) are also found for $CrSi_2P_4$ and $CrGe_2P_4$. The calculated relative energy differences in $MA_2Z_4$ compounds are appreciably smaller than that reported for the single-layer $MoS_2$ (0.27 eV/atom) [81]. This may result in the coexistence of the two phases in experimentally fabricated samples. The $E_{rel}$ values for $MA_2P_4$ monolayers are generally smaller by half than those of $MA_2N_4$. This could be due to the facts that the stability of octahedral coordination increases with increasing the radius of anion.

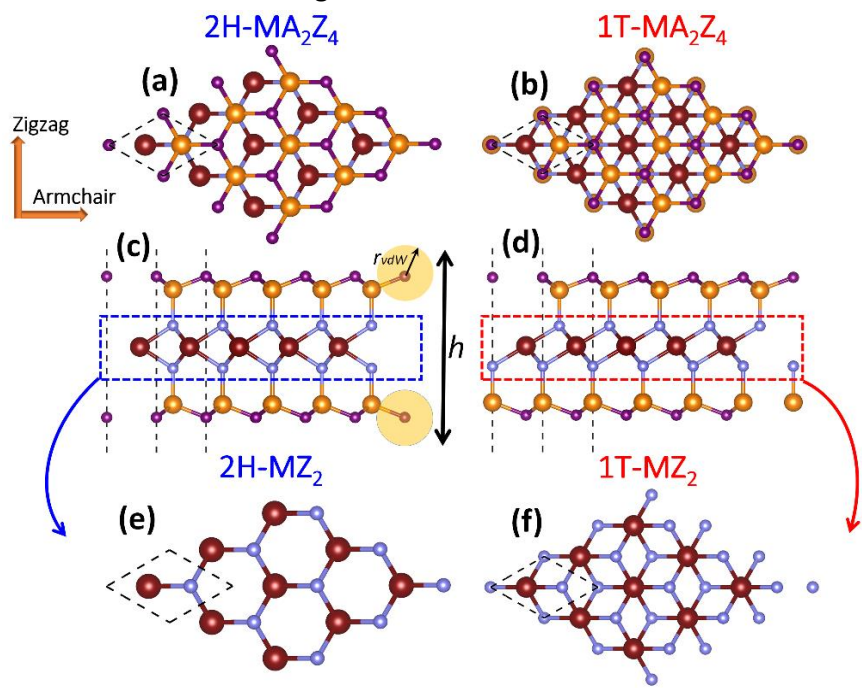

**Fig. 2**, Top and side views of $MA_2Z_4$ monolayers with (a and c)2H and (b and d) 1T phases. The hexagonal primitive lattice is shown with the dashed lines. The monolayer thickness (*h*) is defined as the normal distance between two boundary Z atoms plus their Van der Waals radiuses ($r_{vdW}$). $MZ_2$ motif for (e) 2H and (f) 1T phases are also shown. In this figure light blue and purple colors represent Z (N or P) atoms, wine color M (Cr or Mo or W) atoms, and orange color A (Si or Ge) atoms.

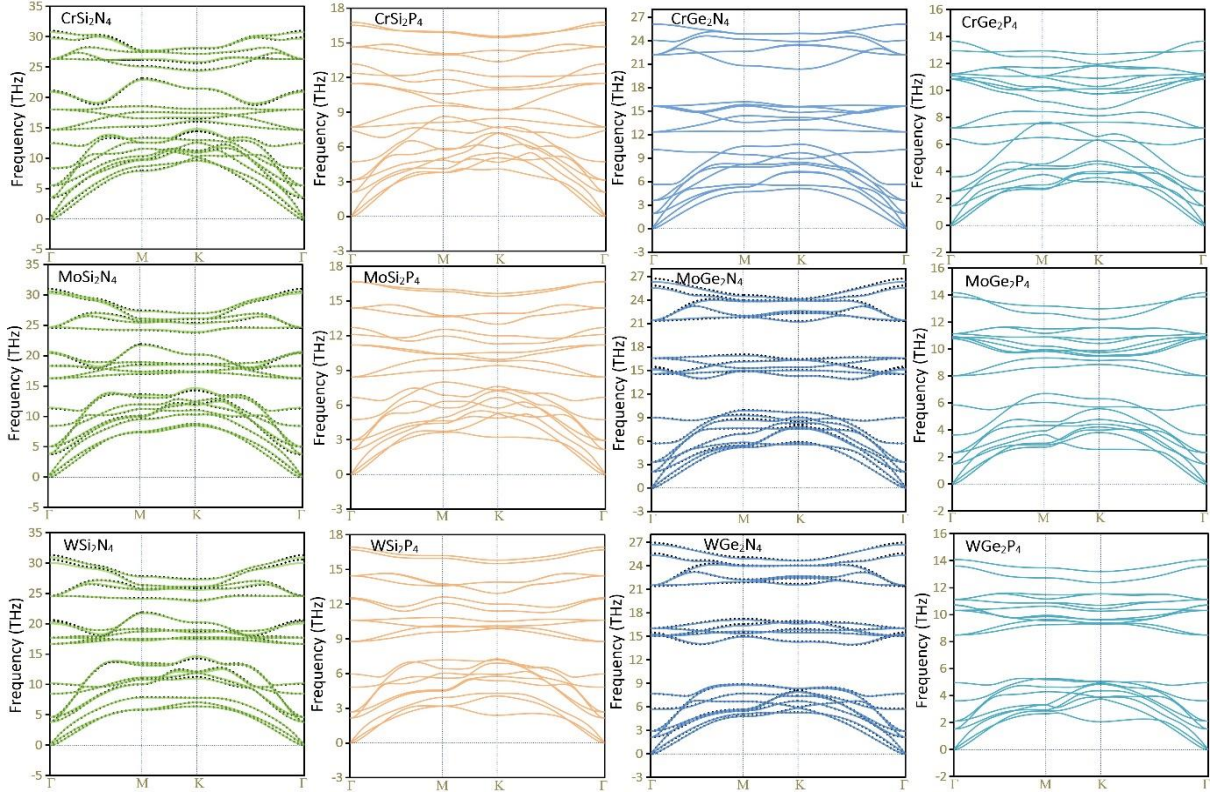

**Fig. 3**, Phonon dispersion relation of 2H-*MA$_2$Z$_4$* monolayer acquired using the MTP method. For several cases the DFPT results are also plotted using the dotted lines.

**Table 1**, Calculated structural, energetic, and electronic properties of *MA$_2$Z$_4$* nanosheets in 2H- and 1T-phase. Lc, h, E$_g$ and E$_{rel}$ are, respectively, hexagonal lattice constant, thickness, band gap and relative energy.

| Lattice | | Lc (Å)[a] | h (Å)[b] | E$_{rel}$ (eV/atom)[c] | Electronic structure[d] | Transition k-points[e] | μ(μ$_B$)[f] |
|---|---|---|---|---|---|---|---|
| CrSi$_2$N$_4$ | 2H | 2.84 | 9.9702 | 0.0 | E$_g^{PBE}$=0.49 eV (I), E$_g^{HSE}$=0.98 eV (I) | (Γ→K) | 0(NM) |
| | 1T | 2.87 | 9.8903 | 0.073 | E$_g^{PBE}$=0 eV (M), E$_g^{HSE}$= 0.07 eV (D) | (Γ →Γ) | 0(AF) |
| CrSi$_2$P$_4$ | 2H | 3.42 | 12.814 | 0.0 | E$_g^{PBE}$=0.28 eV (I), E$_g^{HSE}$=0.64 eV (D) | (K →K) | 0(NM) |
| | 1T | 3.42 | 12.828 | 0.032 | E$_g^{PBE}$=0 eV (M), E$_g^{HSE}$=0 eV (M) | - | 0(AF) |
| CrGe$_2$N$_4$ | 2H | 2.98 | 10.488 | 0.0 | E$_g^{PBE}$=0.49 eV (I), E$_g^{HSE}$= 0.31 eV (I) | (Γ→K) | 0(NM) |
| | 1T | 3.01 | 10.321 | -0.007 | E$_g^{PBE}$=0 eV (M), E$_g^{HSE}$=0 eV (M) | - | 2(FM) |
| CrGe$_2$P$_4$ | 2H | 3.50 | 13.119 | 0.0 | E$_g^{PBE}$=0 eV (M), E$_g^{HSE}$=0 eV (M) | - | 0(AF) |
| | 1T | 3.51 | 13.092 | 0.032 | E$_g^{PBE}$=0 eV (M), E$_g^{HSE}$=0 eV (M) | - | 0(AF) |
| MoSi$_2$N$_4$ | 2H | 2.91 | 10.112 | 0.0 | E$_g^{PBE}$=1.79 eV (I), E$_g^{HSE}$= 2.23 eV (I) | (Γ→K) | 0(NM) |
| | 1T | 2.92 | 10.174 | 0.168 | E$_g^{PBE}$=0 eV (M), E$_g^{HSE}$=0 eV (M) | - | 0(NM) |
| MoSi$_2$P$_4$ | 2H | 3.47 | 12.969 | 0.0 | E$_g^{PBE}$= 0.70 eV (D), E$_g^{HSE}$= 0.99 eV (D) | (K →K) | 0(NM) |
| | 1T | 3.45 | 13.081 | 0.089 | E$_g^{PBE}$= 0.08 eV (I), E$_g^{HSE}$= 0.02 eV (I) | (K →)K→Γ )) | 0(NM) |
| MoGe$_2$N$_4$ | 2H | 3.04 | 10.582 | 0.0 | E$_g^{PBE}$= 0.91 eV (I), E$_g^{HSE}$= 1.27 eV (I) | (K →)K→Γ )) | 0(NM) |
| | 1T | 3.06 | 10.547 | 0.140 | E$_g^{PBE}$=0 eV (M), E$_g^{HSE}$=0 eV (M) | - | 0(NM) |
| MoGe$_2$P$_4$ | 2H | 3.55 | 13.260 | 0.0 | E$_g^{PBE}$= 0.04 eV (I), E$_g^{HSE}$= 0.84 eV (D) | (K →K) | 0(NM) |
| | 1T | 3.54 | 13.342 | 0.084 | E$_g^{PBE}$= 0.04 eV (I), E$_g^{HSE}$= 0.17 eV (D) | (K →K) | 0(NM) |
| WSi$_2$N$_4$ | 2H | 2.91 | 10.114 | 0.0 | E$_g^{PBE}$= 2.07 eV (I), E$_g^{HSE}$= 2.57 eV (I) | (Γ→K) | 0(NM) |
| | 1T | 2.92 | 10.182 | 0.187 | E$_g^{PBE}$=0 eV (M), E$_g^{HSE}$=0 eV (M) | - | 0(NM) |
| WSi$_2$P$_4$ | 2H | 3.48 | 12.960 | 0.0 | E$_g^{PBE}$= 0.53 eV (D), E$_g^{HSE}$= 0.81 eV (D) | (K →K) | 0(NM) |

|  | 1T | 3.46 | 13.061 | 0.089 | $E_g^{PBE}$= 0.04 eV (I), $E_g^{HSE}$= 0.10 eV (I) | (K →)K→Γ )) | 0(NM) |
|---|---|---|---|---|---|---|---|
| WGe$_2$N$_4$ | 2H | 3.04 | 10.586 | 0 | $E_g^{PBE}$= 1.15 eV (I), $E_g^{HSE}$= 1.51 eV (D) | (Γ→K) | 0(NM) |
|  | 1T | 3.06 | 10.551 | 0.153 | $E_g^{PBE}$=0 eV (M), $E_g^{HSE}$=0 eV (M) | - | 0(NM) |
| WGe$_2$P$_4$ | 2H | 3.55 | 13.255 | 0 | $E_g^{PBE}$= 0.48 eV (D), $E_g^{HSE}$= 0.73 eV (D) | (K →K) | 0(NM) |
|  | 1T | 3.55 | 13.324 | 0.082 | $E_g^{PBE}$= 0.10 eV (D), $E_g^{HSE}$= 0.33 eV (D) | (K →K) | 0(NM) |

[a]Hexagonal lattice constants. [b] Effective thickness of a MA$_2$Z$_4$ monolayer is defined as the sum of normal distance between boundary atoms plus the Van der Waals diameter of Z atoms. [c]Relative stability of 2H- and 1T-phases of each compound defined as E$_{rel}$= E$_{tot}$(2H-MA$_2$Z$_4$) – E$_{tot}$(1T-MA$_2$Z$_4$). [d]PBE and HSE06 calculated band gaps. Abbreviations "I" and "D" indicate indirect and direct gap semiconducting character, respectively, and "M" indicate metallic character. [e]band gap transition k-points of HSE06 results. [f]Total magnetic moment per rectangular unitcell (two M atoms). Abbreviations "NM", "FM", and "AF" indicate nonmagnetic, ferromagnetic, and antiferromagnetic ground states spin order in MA$_2$Z$_4$ nanosheets, respectively.

We next investigate the electronic properties of MA$_2$Z$_4$ monolayer in 2H and 1T phases via calculating their electronic band structures along the high symmetry direction of the Brillouin zone. The HSE06 and PBE band structures of 2H-MA$_2$Z$_4$ monolayers are depicted in Fig. 3 and those for the corresponding 1T phases are shown in Fig. S2. The predicted band gap values, nature of band gaps, and HSE06 transition k-points are also listed in Table 1. It is conspicuous that this family exhibits diverse electronic properties and depending on their chemical compositions they show metallic to moderate band gap semiconducting nature. Several observations can be concluded from the band structure results, as follows: (1) All 2H monolayers are semiconducting except for 2H-CrGe$_2$P$_4$, which is found to be a AF metal. The predicted band gap values range from 0.31 eV for 2H-CrGe$_2$N$_4$ to 2.57 eV for 2H-WSi$_2$N$_4$. (2) The valence band maximum (VBM) in 2H-MA$_2$N$_4$ and 2H-MA$_2$P$_4$ is located at Γ and K points, respectively, while conduction band minimum (CBM) for all 2H semiconductors always occurs at K point. As a consequence, 2H-MA$_2$N$_4$ are indirect band gap materials (Γ→K), while 2H-MA$_2$P$_4$ are direct ((K →K)) gap semiconductors. Predicted band gap values and transition *k*-points for the experimentally synthesized 2H-CrSi$_2$N$_4$ (2.23 eV, Γ→K) are in very good agreement with those (2.297 eV, Γ→K) reported in the original study [47]. Interestingly, for 2H-CrSi$_2$N$_4$ and 2H-MoSi$_2$N$_4$ the gap at point K is only 0.06 eV larger than the indirect gap, indicating that these two monolayers can be considered as quasi-direct gap semiconductors. These observations are highly promising for potential applications in electronic and optoelectronic devices. (3) 2H-MA$_2$N$_4$ show greater band gaps than corresponding MA$_2$P$_4$ counterparts. (4) Generally 1T monolayers exhibit appreciably smaller band gaps than their corresponding 2H lattices. More precisely, most of 1T monolayers even exhibit metallic character. In this regard the only dynamically stable lattice of 1T-CrGe$_2$N$_4$, is a half metal, and the others are semiconductors with band gaps smaller than 0.33 eV (find Table 1 and Fig. S2). A very similar band gap trend is also observed for 1T and 2H phases of single layer transition metal dichalcogenides [82].

To better understand the nature of band edge states and also to rationalize some of the observed trends, we calculated the charge density distribution at VB(Γ), VB(K) and CB(K) of each 2H-MA$_2$Z$_4$ monolayer, shown in Fig. S3. Our charge density analysis reveals that for all monolayers, VB(Γ) is mainly constructed by M-$d_{z^2}$ orbital with minor contribution by Z-p$_z$ and A-p$_z$ orbitals, representing a σ(M-M) state hybridized with σ(A-Z) states. However, VB(K) is predominantly distributed over MZ$_2$ layer and it represents σ(M-M) state hybridized with σ(M-Z) states derived from M-($d_{x^2-y^2}$, $d_{xy}$) and Z-(p$_x$, p$_y$, p$_z$) orbitals. The CB(K) is solely derived from M-(s,$d_{z^2}$) orbitals. Our analysis indicates the key role of metal-metal interaction in constructing band edge states of these materials. Owing to the strong interaction between d orbitals of M atoms, the VB and CB are both highly dispersed, implying small effective masses and high charge carrier mobilities. From Fig. 4, it is evident that 2H-MA$_2$N$_4$ monolayers exhibit greater VB and CB dispersions and consequently smaller electron and hole effective masses than those of 2H-MA$_2$P$_4$ monolayers. This observation is due to the smaller lattice constants of 2H-MA$_2$N$_4$ monolayers which results in more effective metal-metal interaction (See Fig. S3) and consequently higher band dispersion. Here, we recall our observation above that 2H-MA$_2$N$_4$ are indirect gap semiconductors (Γ→K), whereas 2H-MA$_2$P$_4$ monolayer are direct gap semiconductors (K →K). By moving from 2H-MA$_2$P$_4$ to 2H-MA$_2$N$_4$ both VB(Γ) and VB(K) are stabilized due to the stronger metal-metal interaction in 2H-MA$_2$N$_4$. However, because the planar $d_{x^2-y^2}$,$d_{xy}$ orbitals in VB(K), higher d-d orbital overlap occur as compared to the $d_{z^2}$ orbitals in VB(Γ), and VB(K) state becomes more deeply stabilized than VB(Γ) state and the material turns into an indirect band gap semiconductor.

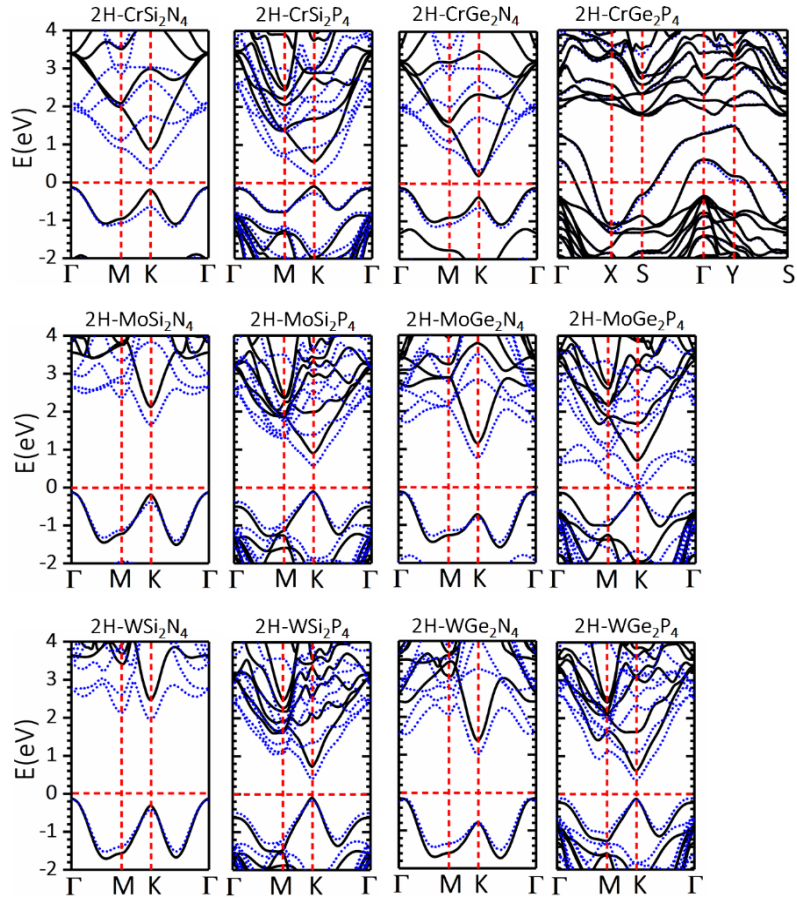

**Fig. 4**, HSE06 and PBE band structures for 2H-MA$_2$Z$_4$ monolayers in their magnetic ground states. Black solid and blue dotted lines represent HSE06 and PBE band structures, respectively. The Fermi level is set to 0 eV.

After examining the stability and electronic features, we now explore the photocatalytic activity of *MA$_2$Z$_4$* monolayers for water splitting reaction from a thermodynamic point of view.

In order to achieve photocatalytic water splitting using a single semiconductor, a material must possess a band gap in the range of 1.23-3 eV to be able to harvest visible light. According to Table 1, only four of $MA_2Z_4$ monolayers satisfy this criterion, namely, 2H-MoSi$_2$N$_4$, 2H-MoGe$_2$N$_4$, 2H-WSi$_2$N$_4$, and 2H-WGe$_2$N$_4$ with band gap values of 2.23, 1.27, 2.57, and 1.51 eV, respectively. We recall that 2H-MoSi$_2$N$_4$ is a quasi-direct gap material and the other three lattices are indirect gap semiconductors. In Fig. 5, the band edge position of these four monolayer calculated using HSE06 are shown with respect to the vacuum level. The standard potentials for the two half reactions of decomposition of water (2H$^+$(aq) + 2e$^-$ → H$_2$(g), H$_2$O→ 2H$^+$ + 1/2O$_2$) are also drawn at pH= 0 and 7. For hydrogen evolution reaction (HER), the standard reduction potential at pH = 0 is -4.44 eV, while that of the oxygen evolution reaction (OER) is by 1.23 eV lower than the HER level (-5.67 eV). In addition, the standard potential of both HER and OER sensitively vary with pH according to the Nernst equations of $E^{red}$ (H$^+$/H$_2$) = −4.44 eV + pH × 0.059 eV and $E^{ox}$ (H$_2$O/O$_2$) = −5.67 eV + pH × 0.059 eV. Besides having a large band gap (> 1.23eV), the band edge positions of a photocatalyst candidate must straddle the standard redox potentials of water (SRPW). In Fig. 5a, it can be clearly seen that band edge positions of 2H-MoSi$_2$N$_4$ straddle the SRPW in both highly acidic (pH=0) and neutral (pH=7) conditions. However, those of 2H-WSi$_2$ and 2H-WGe$_2$N$_4$ straddle the SRPW only in highly acidic and neutral conditions, respectively. We predicted that the CBM of 2H-MoGe$_2$N$_4$ occurs at appreciably lower energies than standard potential of OER reaction in acidic condition and therefore it cannot function as photocatalyst for water splitting. This was expectable due to a relatively small band gap in 2H-MoGe$_2$N$_4$ monolayer.

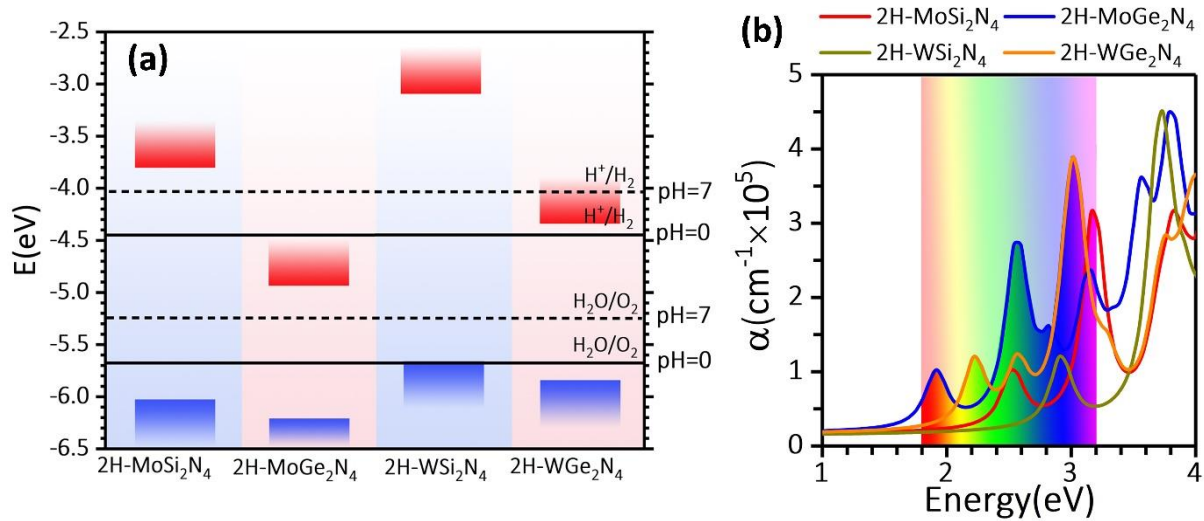

**Fig. 5,** (a) Band edge positions with respect to the vacuum level for those 2H-$MA_2Z_4$ monolayers with band gap larger than 1.23 eV. For comparison, the potential of the two half-reactions of HER (H$^+$/H$_2$) and OER (H$_2$O/O$_2$) are also shown at pH=0 (solid horizontal lines) and pH= 7 (dotted horizontal lines). (b) Optical absorption coefficient obtained using HSE06 functional for 2H-WGe$_2$N$_4$, WSi$_2$N$_4$, MoGe$_2$N$_4$ and MoSi$_2$N$_4$ monolayers. The light incident is along the out-of-plane direction and isotropically polarized along the in-plane directions. The energy range of visible light spectrum is also shown.

In addition to the two criteria investigated above, a photocatalyst material should also yield high electron and hole mobilities, so that the photogenerated electron and holes can participate in charge transfer process before they recombine. We estimated the electron and hole mobilities of the four monolayers along armchair and zigzag directions using the deformation potential theory (DPT) in combination with the effective mass approximation. It is found that both electron and hole mobilities are almost isotropic along armchair and zigzag directions. Our predicted electron and hole mobilities are, respectively, 200 and 1100 cm$^2$V$^{-1}$s$^{-1}$ for 2H-MoSi$_2$N$_4$, 490 and 2190 cm$^2$V$^{-1}$s$^{-1}$ for 2H-MoGe$_2$N$_4$, 320 and 2026 cm$^2$V$^{-1}$s$^{-1}$ for 2H-WSi$_2$N$_4$, 690 and 2490 cm$^2$V$^{-1}$s$^{-1}$ for 2H-WGe$_2$N$_4$. Our calculated mobilities for 2H-MoSi$_2$N$_4$ are in very good agreement with those reported in a previous work (~270 and 1200 cm$^2$V$^{-1}$s$^{-1}$ for electron and hole mobilities, respectively [47]. The calculated mobilities are much larger than those theoretically estimated for MoS$_2$ monolayer (72.16 and 200.52 cm$^2$V$^{-1}$s$^{-1}$ for electron and hole mobilities, respectively) [83]. In all cases, the electron mobility is found to be a few times larger than the hole mobility. The large difference between electron and hole mobilities may enhance electron-hole separation efficiency and photocatalytic activity. A photocatalyst is also required to exhibit absorption in visible light. We thus examine the optical absorption of these monolayers by computing their dielectric functions and absorption coefficients ($\alpha(\omega)$) using the HSE06 functional. Fig. 5b depicts absorption spectrum in response to the light incident along the out-of-plane direction and polarized along the in-plane directions (find Fig. S4 for more details). Considered monolayers are all found to exhibit isotropic absorption, due to their highly symmetrical lattice. Although 2H-MoSi$_2$N$_4$ is a quasi-direct gap semiconductor and the three others are indirect gap, it is evident that the four monolayers absorb light in the visible range. The calculated absorption coefficients (10$^5$ cm$^{-1}$) are comparable to those of perovskites, which are known to be highly efficient for solar cells.

We next examine the mechanical responses of 2H-*MA$_2$Z$_4$* monolayers according to the uniaxial stress-strain curves. In these simulations the stresses along the two perpendicular directions of the loading are ensured to stay small. Along the normal direction of the monolayers this objective is automatically achieved upon the geometry optimization due to the contact with vacuum. For the other planar direction, the stress is reached to a negligible value by altering the periodic box size [84]. Mechanical responses are evaluated along the armchair and zigzag (as shown in Fig. 2) directions to assess the anisotropy. The acquired stress-strain relations of *MA$_2$Z$_4$* monolayers for the loading along the armchair and zigzag directions are illustrated in Fig. 6. The key mechanical properties including the elastic modulus and tensile strengths are also summarized in Table 2. As it is clear, for all the considered monolayers loaded along the armchair and zigzag directions, the initial linear parts of the stress-strain relation coincide, confirming the isotropic elasticity. Moreover, by increasing the strain level and deviating from the initial linear response, the samples loaded along the armchair direction start to exhibit higher stresses than those along the zigzag direction.

Acquired results suggest that depending on the terminating elements of N or P atoms, *MA$_2$Z$_4$* nanosheets show different behaviors. For the MA$_2$N$_4$ monolayers loaded along the armchair direction, the maximum tensile strengths happen at lower strains than the zigzag direction. For these structures, while the structures show higher stretchability along the zigzag direction, the tensile strengths along the both loading directions are close. In MA$_2$P$_4$ nanosheets, the maximum tensile strengths however occur at closer strain levels for the loading along armchair and zigzag. This way, $MA_2P_4$ nanosheets show noticeably higher tensile strengths along the armchair than the zigzag direction. It is found that by increasing the atomic weight of the core atoms in $MA_2Z_4$ nanosheets, the elastic modulus increases slightly. The elastic modulus of $MoSi_2N_4$ is found to be 487 GPa, showing an excellent agreement with experimentally measured value of 491.4 ± 139.1 GPa [47]. For the tensile strength, $WA_2Z_4$ nanosheets exhibit higher values than $MoA_2Z_4$ and $CrA_2Z_4$ counterparts. The tensile strength of $MoSi_2N_4$ is predicted to be 55.4-57.8 GPa, which is within the range of experimentally reported value of 65.8 ± 18.3 GPa [47]. From the acquired results it is clear that the mechanical properties are mainly affected by the type of terminating atoms, and such that lattices composed of N atoms show substantially higher mechanical characteristics than those terminated with P atoms. It is conspicuous that while the core atoms slightly affect the mechanical properties, the increasing of the atomic weight of A and Z atoms in $MA_2Z_4$ nanosheets remarkably suppress the both elastic modulus and tensile strength. From the structural point of view, in $MA_2Z_4$ nanosheets only M-Z and A-Z bonds are partially oriented along the loading direction. These bonds are thus directly involved in the deformation and can result in a dominant effect on the overall mechanical properties. From the general chemistry we know that chemical bonds formed with N atoms are appreciably stronger than those made with P atoms and moreover Si-N bond is stronger than Ge-N one, resulting in the maximal elastic modulus for $MSi_2N_4$ monolayers. It is also noticeable that the $MSi_2N_4$ lattices normally exhibit higher maximal stretchability. It can be thus concluded that in order to reach maximal mechanical characteristics, $MSi_2N_4$ nanosheets show superiority and the type of M core atom is not expected to result in substantial changes.

Table 2, Summarized elastic modulus (E) and tensile strength (TS) of 2H-$MA_2Z_4$ monolayers along the armchair (Arm.) and zigzag (Zig.) directions. Room temperature lattice thermal conductivity (K) and contribution of ZA, TA and LA acoustic and optical modes on the overall conductivity.

| Lattice | E (GPa) | $TS^{Zig.}$ (GPa) | $TS^{Arm.}$ (GPa) | K (W/mK) | Contribution | | | |
|---|---|---|---|---|---|---|---|---|
| | | | | | ZA | TA | LA | Optical |
| $CrSi_2N_4$ | 468 | 57.8 | 55.4 | 332 | 0.24 | 0.28 | 0.19 | 0.29 |
| $MoSi_2N_4$ | 487 | 56.7 | 53.3 | 439 | 0.21 | 0.26 | 0.32 | 0.21 |
| $WSi_2N_4$ | 506 | 59.2 | 55.5 | 503 | 0.25 | 0.22 | 0.32 | 0.21 |
| $CrGe_2N_4$ | 340 | 38.7 | 38.1 | 198 | 0.21 | 0.26 | 0.33 | 0.20 |
| $MoGe_2N_4$ | 362 | 40.3 | 42.1 | 286 | 0.19 | 0.26 | 0.32 | 0.24 |
| $WGe_2N_4$ | 384 | 42.6 | 44.5 | 322 | 0.24 | 0.21 | 0.23 | 0.32 |
| $CrSi_2P_4$ | 154 | 18.5 | 21.2 | 120 | 0.23 | 0.25 | 0.32 | 0.20 |
| $MoSi_2P_4$ | 159 | 17.5 | 21.2 | 122 | 0.23 | 0.20 | 0.33 | 0.24 |
| $WSi_2P_4$ | 167 | 18.8 | 22.0 | 129 | 0.22 | 0.22 | 0.31 | 0.25 |
| $CrGe_2P_4$ | 135 | 15.2 | 17.7 | 63 | 0.21 | 0.25 | 0.33 | 0.21 |
| $MoGe_2P_4$ | 139 | 15.3 | 18.4 | 63 | 0.17 | 0.20 | 0.37 | 0.26 |
| $WGe_2P_4$ | 145 | 16.5 | 19.3 | 64 | 0.17 | 0.20 | 0.29 | 0.34 |

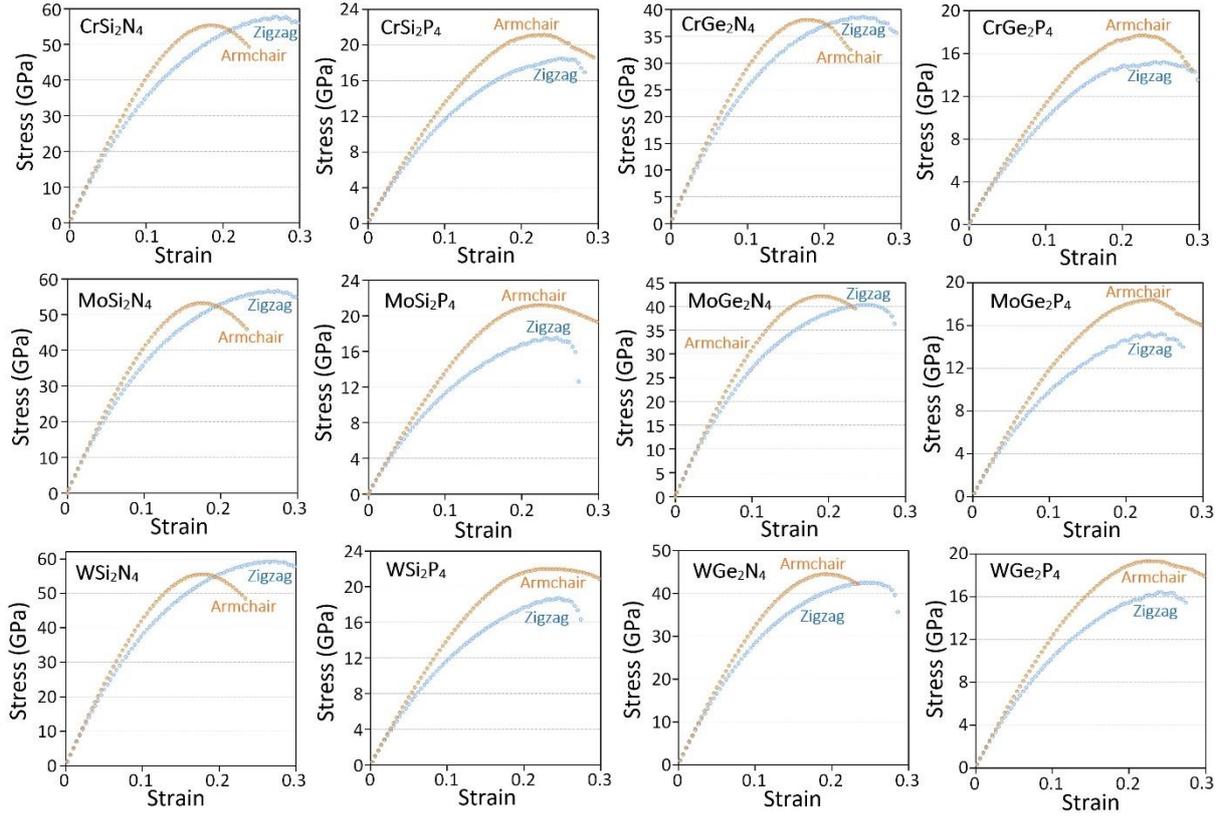

**Fig. 6**, Uniaxial stress-strain responses of 2H-*MA₂Z₄* monolayers for the loading along the armchair and zigzag directions.

We next evaluate the lattice thermal conductivity of 2H-*MA₂Z₄* monolayers at the room temperature using the full iterative solutions of the Boltzmann transport equation. For the solution of lattice thermal conductivity, the anharmonic interatomic force constants for every composition is obtained by computing the interatomic forces for 784 systems, each one consisting of 175 atoms. In Fig. 3 the phonon dispersion relations of 2H-*MA₂Z₄* monolayers are depicted. For few cases the MTP-based results are compared with those by DFPT method, which show excellent agreements, in accordance with our earlier study results [62]. In principle, exhibiting a wider or narrower dispersion for a particular band, particularly for the acoustic modes, reveal faster or slower group velocity, respectively. Moreover, the phonon bands with narrow dispersions crossing each other may increase scattering and result in lower thermal conductivity. By considering the range of frequencies for the phonons, it is clear that the type of the core atom shows marginal effects. In contrast, by changing the type of terminating atom the phonon dispersions and the maximum range of frequencies change considerably. In general, MA$_2$N$_4$ monolayers show higher frequencies than MA$_2$P$_4$ counterparts. Similarly, MSi$_2$Z$_4$ lattices show wider phonon dispersions than MGe$_2$Z$_4$ lattices. As mentioned earlier, the compression of phonon bands may result in lower group velocity and higher scattering and subsequently lower thermal conductivity. Therefore, it is expected that the lattice thermal conductivity decreases by going from MSi$_2$N$_4$ to MGe$_2$N$_4$, MSi$_2$P$_4$ and MGe$_2$P$_4$. In Table 2 we also summarized the values of room temperature lattice thermal conductivity for 2H MA$_2$Z$_4$ monolayers, which confirms this trend. Worthy to mention that the studied monolayers are all found to show isotropic lattice thermal conductivity. In Fig. S5, we compare the phonons group velocity of MoSi$_2$N$_4$, MoGe$_2$N$_4$, MoSi$_2$P$_4$ and MoGe$_2$P$_4$ monolayers, respectively, which reveal noticeable suppressions of phonon group velocities and thus explaining the observed decreasing trend in thermal conductivity.

As we discussed earlier, by increasing the weight of the core atom the elastic modulus of *MA$_2$Z$_4$* nanosheets increases. Acquired results reveal that by increasing the atomic weight of the core atom the thermal conductivity of *MA$_2$Z$_4$* monolayers increases, which is consistent with the classical theory that expects a higher thermal conductivity for stiffer systems. The higher thermal conductivity of WA$_2$Z$_4$ than MoA$_2$Z$_4$ is also consistent with the reported trend for the WS$_2$ and MoS$_2$ monolayers [85]. In Table 2, we also compare the contribution of acoustic and optical modes on the overall lattice thermal conductivity. It is conspicuous that the acoustic modes are the main heat carriers in the studied systems, by average yielding 75% of the total phononic thermal conductivity. As an example, in Fig. S6 we compare the phonon lifetime of WSi$_2$N$_4$ and MoSi$_2$N$_4$ monolayers. Our results generally show noticeably higher lifetimes for the acoustic modes (low range of frequencies) in WSi$_2$N$_4$ in comparison with the MoSi$_2$N$_4$ counterpart. Since the acoustic modes are the main heat carriers in these systems, it can be concluded that the higher scattering rates of acoustic modes in MoSi$_2$N$_4$ monolayer decrease the phonons' lifetime and result in a lower thermal conductivity than WSi$_2$N$_4$ counterpart. In Fig. S7 we also compare the cumulative thermal conductivity as a function of phonons mean free path, which reveals that the lattice thermal conductivity usually converges for the lengths between 20 to 100 μm. Our results reveal a consistent trend for the elastic modulus and lattice thermal conductivity, such that a *MA$_2$Z$_4$* monolayer with a higher elastic modulus shows a higher thermal conductivity as well. It can be thus suggested that in order to reach a noticeably high thermal conductivity, W core atom is more favorable than Mo and Cr counterparts. The thermal conductivity of WSi$_2$N$_4$ and MoSi$_2$N$_4$ monolayers are found to be 503 and 439 W/mK, which are by several folds higher than WS$_2$ and MoS$_2$ monolayers [85]. The remarkably high thermal conductivity of MSi$_2$N$_4$ nanosheets is highly promising for the thermal management systems. Worth noting that while the thermal conductivity of these novel 2D systems are yet by around 8 folds lower than that of the graphene, but their semiconducting electronic nature is more appealing for the thermal management in nanoelectronics and Li-ion batteries than the semimetal character of graphene.

Finally, we explore the piezoelectric and flexoelectric responses of this novel class of 2D materials. Fig.7a and 7b show the variation of polarization $P_y$ with strain $\epsilon_{yy}$ in MA$_2$Z$_4$ monolayers. We first consider MoSi$_2$N$_4$ and MoSi$_2$P$_4$ monolayers for exploring the variation in polarization with deformation. As it is clear, a linear variation between $P_y$ and $\epsilon_{yy}$ exists. The atomic deformations using the displacement filed $u_y$ helps to dismiss the flexoelectric part of polarization from Eq. 3. In the case of tensile deformation, Fig. 7 shows nearly constant variation for strain component $\epsilon_{yy}$ along the $y$ direction. This confirms that strain gradient is zero and removes the flexoelectric contribution during the tensile deformation $\left(\mu_{\alpha\beta\gamma\delta} \frac{\partial \epsilon_{\beta\gamma}}{\partial r_\delta} = 0\right)$. Note that the strain values represented in Fig. 7 are the averaged values of the atomic strain over 23 equally spaced bins along the $y$ direction. The wavy nature of the strain variation is due to the uncontrollable out-of-plane thermal fluctuations. The slope of variation between $P_y$ and $\epsilon_{yy}$ is the piezoelectric coefficient $d_{yyy}$, which is 2.293 nC/m for MoSi$_2$N$_4$ and 0.890 nC/m for MoSi$_2$P$_4$. In general, the dipole moment of an atom depends on the effective atomic polarizability and total electric filed induced by the charges and dipoles. The unit cell polarizability of MoSi$_2$N$_4$ and MoSi$_2$P$_4$ are 36.372 and 80.274 Å$^3$, respectively. Fig. 8a and 8b show the atomic configuration for MoSi$_2$N$_4$ system at zero strain (stress-free) and at a strain of 0.01, respectively. The total electric field difference $\Delta\left(E_y^q + E_y^p\right)$ between these configurations is 182.721 V/Å in MoSi$_2$N$_4$ and 37.212 V/Å in MoSi$_2$P$_4$ monolayers. The ratio of polarizability $\Delta\left(E_y^q + E_y^p\right)$ between MoSi$_2$N$_4$ and MoSi$_2$P$_4$ is about 2.225, which is in a close match with the ratio of piezoelectric coefficients.

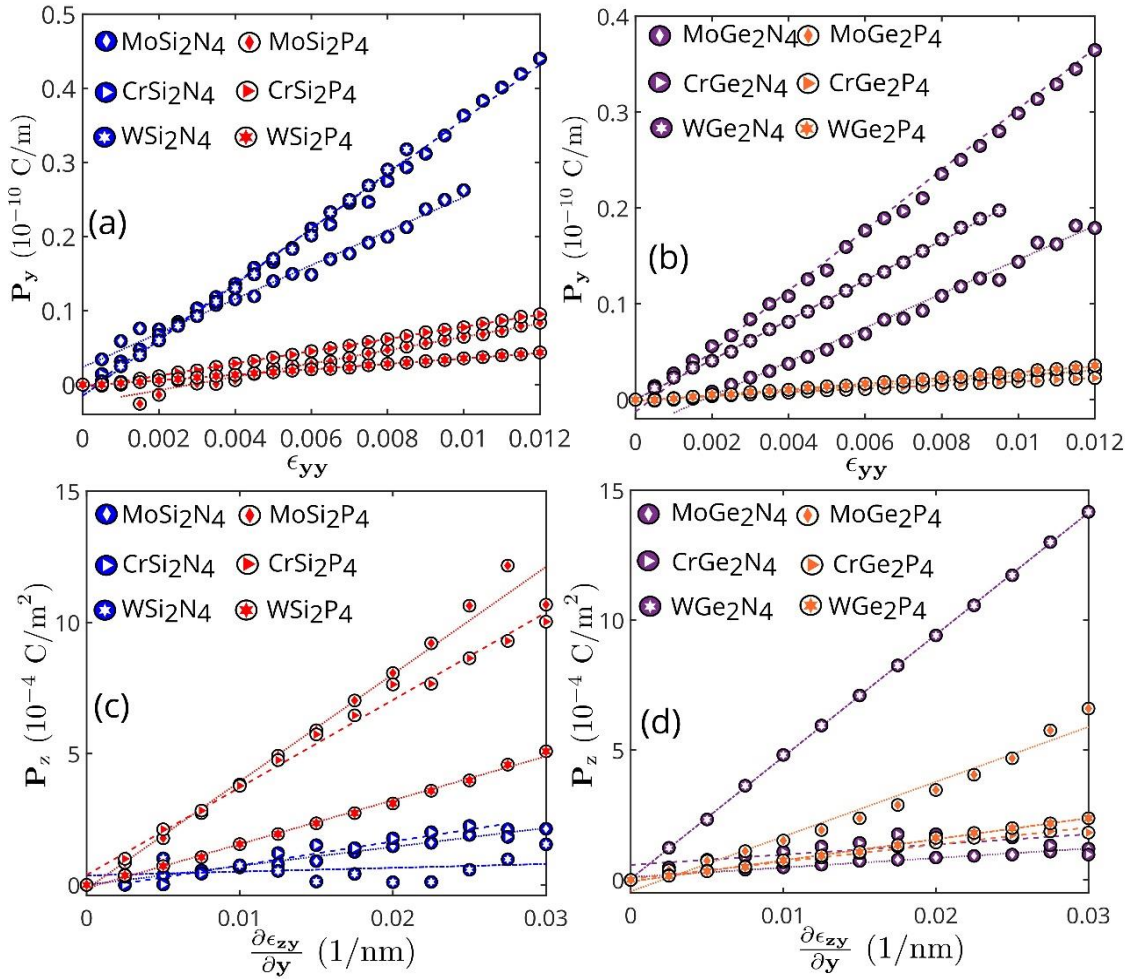

**Fig. 7**, The variation of polarization of $P_y$ with strain $\epsilon_{yy}$ for (a) MSi$_2$Z$_4$ and (b) MGe$_2$Z$_4$ systems. The bending induced polarization $P_z$ with strain gradient $\frac{\partial \epsilon_{zy}}{\partial y}$ for (c) MSi$_2$Z$_4$ and (d) MGe$_2$Z$_4$ systems. The solid lines indicate the linear fitting to the respective simulation data.

Fig. 8c shows the dipolar distribution for MoSi$_2$N$_4$ nanosheets under the deformed configuration. The dipole moment $(p_y)$ is nearly zero for all the N (B, D, F, G, I and K labels) atoms and 0.012 for Mo atom (label A). Whereas the Si atoms (C and J labels) left to Mo atom exhibit 0.224 $e$Å as $p_y$ and right (E and H labels) to Mo atom show 0.125 $e$Å. The change in dipole moment between Si atoms is due to the differences in the elongation in the bond C-D and D-E. For undeformed MoSi$_2$N$_4$ (Fig. 8a), C-D and D-E bond lengths are about 1.755 Å. The symmetric bond lengths induce equal and opposite local electric fields, which does not help to generate the dipole moment. Because of the tensile deformation, the C-D bond stretches to 1.794 Å and D-E bond compresses to 1.696 Å. This difference in bond lengths breaks the electric field symmetry and induce dipole moments according to the local electric field strength. Similarly, the changes in bond lengths connected to Mo atom helps to induce the observed smaller dipole moment. The changes in electric fields implies the importance of $\pi - \sigma$ and $\sigma - \sigma$ interactions originated across the valence and bonding electrons in generating the dipole moments. The electron affinity of N is higher compared to Mo and Si atoms, which means N atoms are reluctant to change their charge state. The difference in charge state for atom labels B, D and F between strained and unstrained configurations is about 0.002$e$. The smaller change in charge represent that there exists a weak transfer of charge to atom N. This reduce the $\pi - \sigma$ or $\sigma - \sigma$ interactions in N and induce low dipole moments. Whereas, for atom label C the difference in charge is about 0.0426$e$, which shows an easy transfer of charge to Si atom and thus resulting in high dipole moment via enhanced $\pi$ and $\sigma$ interactions. The change in charge state for atom label E is about 0.01$e$, which results into a low dipole moment. We note the similar observations for atoms below the Mo atom in Fig. 8c. Finally, the total dipole moment of MoSi$_2$N$_4$ is predominantly from the Si atoms. The sum over these dipole moments across all the unitcells leads to the observed high dipole moment and polarization for MoSi$_2$N$_4$. We consider MoSi$_2$P$_4$ atomic configuration at the same strain state (as shown in Fig. 8c). Here, there is an improvement of the dipole moment of P atoms over N atoms in MoSi$_2$N$_4$. The low electron affinity of P over N helps to change their charge state easily and allow for increasing the $\pi - \sigma$ and $\sigma - \sigma$ interactions, which helps in the form of increasing the $E_y^q$ and $E_y^p$ parts of the total electric field. However, the total dipole moment or polarization of MoSi$_2$P$_4$ is much lower than MoSi$_2$N$_4$ due to the oppositely induced dipole moments of Si and P atoms. Besides, Mo atoms also reduce the total dipole moment due to the increased electric field effect from the P atoms. It can be thus concluded that due to the chemical nature of N and P atoms, MoSi$_2$N$_4$ exhibits a large dipole moment compared to MoSi$_2$P$_4$ under tensile deformation.

CrSi$_2$Z$_4$ and WSi$_2$Z$_4$ nanosheets also follow the above observations during the comparison between N and P substitution. The piezo coefficient for CrSi$_2$N$_4$ and WSi$_2$N$_4$ are 3.687 and 3.760 nC/m, respectively and $d_{yyy}$ for CrSi$_2$N$_4$ is by a factor of 1.6 higher than that of the MoSi$_2$N$_4$. In CrSi$_2$N$_4$, the ratio of polarizability (from Table 3) times $\Delta(E_y^q + E_y^p)$ over MoSi$_2$N$_4$ is about 1.345, which is nearer to the observed increase in piezoelectric coefficients. Here Si atoms left to Cr obtain a dipole moment of 0.595 and right to Cr show 0.367 and these values are higher than those observed in MoSi$_2$N$_4$. Cr and N atoms show further lower dipole moments than in MoSi$_2$N$_4$ due to their high electron affinities. Thus the given mechanical deformation is efficiently used only by the Si atoms to change their charge state and induce larger $\pi - \sigma$ and $\sigma - \sigma$ interactions, which enhances the local electric fields and results in higher dipole moments. For the case of WSi$_2$N$_4$, W gets a dipole moment of 0.031eÅ, which is by 2.58 times higher than the dipole moment of atom Mo in MoSi$_2$N$_4$. The lower electron affinity of atom W helps to raise its dipole moment over Mo and Cr. The low electron affinity helps to change the charge state easily and develop required interactions to induce dipole moment. The Si atoms left to W have 0.216 eÅ and right atoms show 0.114 eÅ as dipole moments, which are reduced by a small amount when compared to Si in MoSi$_2$N$_4$. The increased contribution from W helps to increase the total polarization of the WSi$_2$N$_4$ system. In total, the ratio of polarizability times the change in electric field across configurations of WSi$_2$N$_4$ and MoSi$_2$N$_4$ is about 1.640. This ratio is in close agreement with the proportion between their piezoelectric coefficients. Identical observations appear in enhancing the piezoelectric coefficients for materials involved with Ge (find Fig. 7b).

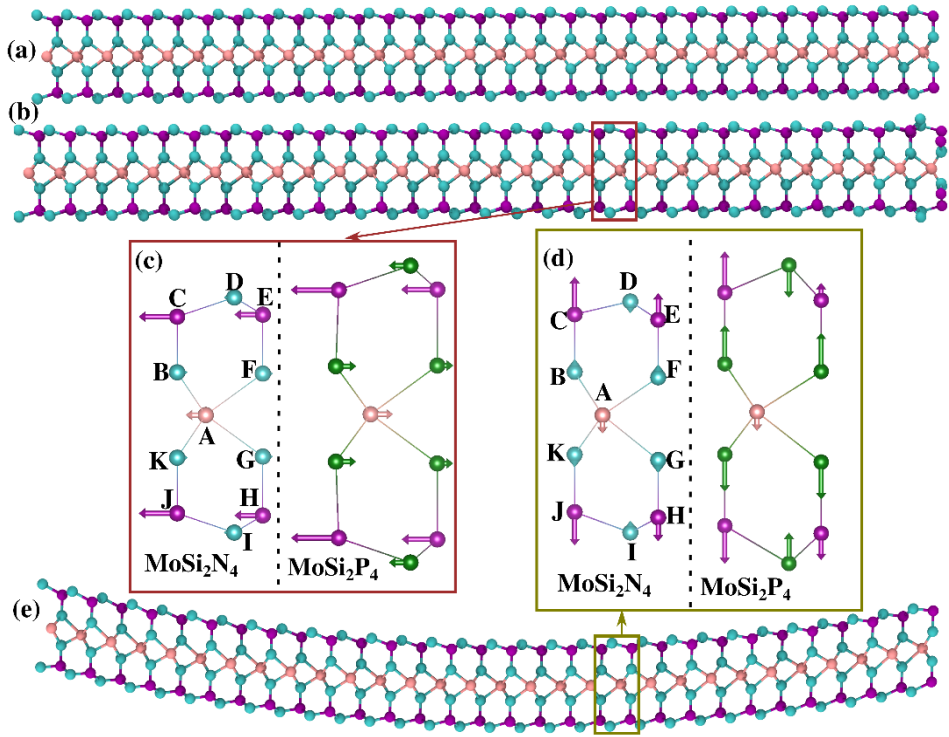

**Fig. 8**, Atomic configuration for MoSi$_2$N$_4$ system at (a) undeformed state and (b) tensile deformed state with strain 0.01. (e) represents the atomic configuration during the bending deformation at strain gradient of 0.3 Å$^{-1}$. (c) and (d) shows the selected unitcell atoms in the tensile and bending deformed configurations. The arrows indicate the respective dipole moments. MoSi$_2$N$_4$ and MoSi$_2$P$_4$ unitcells are included in (c) and (d) panels, respectively.

Fig. 7c shows the variation of polarization $P_z$ due to the bending deformation by increasing the strain gradients. The values of $P_z$ are calculated similar to $P_y$ and divided with thickness mentioned in Table 1 to match with the units of flexoelectric coefficients. Here also, Fig. 1 also confirms the linear variation for the strain component $\epsilon_{zy}$ with respect to bins in y-direction during the bending deformation. The linear variation further cancels the total strain $\epsilon_{zy}$ and maintains the piezoelectric contribution $d_{\alpha\beta\gamma}\epsilon_{\beta\gamma}$ as zero for bending deformation. However, the slope of the $\epsilon_{zy}$ component is highly in match with the given strain gradient $\frac{1}{2}K_b$ at the given load step. MoSi$_2$N$_4$ atomic configuration in Fig. 8e is at $\frac{1}{2}K_b$ or $\frac{\partial \epsilon_{zy}}{\partial y}$ equal to 0.3 Å$^{-1}$. Note that the training dataset used to generate MTP does not contain any information related to bending deformation. However, the developed MTP can predict the bending deformation. The atomic configuration in Fig. 8e deformed uniformly in the left and right parts of the sheet. We consider the same unitcell used in tensile deformation for dipole moment analysis in bending case. For MoSi$_2$N$_4$, the N atoms have low dipole moments due to the weak $\pi - \sigma$ interactions. Here top and bottom Si atoms are having dipole moments opposite to each other. The asymmetry in bond and angle values associated with Si atoms leads to change in the dipole moments via the local electric fields. During bending, the bond C-D compressed to 1.726 Å and D-E stretched to 1.771 Å, respectively. However, the change in bond angle ∠BCD is higher than bond angle ∠FED. Since the angle change induces a shift in buckling height between successive atoms, which helps to achieve a different hybridization state via the $\pi - \sigma$ interactions. These changes in bond length and angles lead to the higher dipole moment for the C labeled-atom than the E one. For the case of tensile deformation atom labels, C and J have identical dipole moments. In contrast to tensile case, the dipole moments for atom label J is lower than dipole moment of atom C because of the changes in local atomic configuration in response to the upward bending deformation. Between atom labels H and E also similar changes in the dipole moments are observed. In total, these changes in dipole moments added up, to produce polarization. The linear variation between $P_z$ and strain gradient for MoSi$_2$N$_4$ gives a flexoelectric coefficient of 0.007 nC/m. In case of MoSi$_2$P$_4$, Si atoms are performing in the similar manner as MoSi$_2$N$_4$. P atoms (B, F, G and K) produce dipole moments in the direction of nearest bonded Si atoms. Interestingly, atoms D and I (type P) produce significant contribution to the total dipole moment because of the differences in angle and bond lengths, which induce different local electric fields for these atoms and producing non vanishing dipole moments. This helps to increase the total dipole moment of MoSi$_2$P$_4$ unitcell to 6.87 times higher than MoSi$_2$P$_4$ counterpart. The rise in dipole moment reflected in the increased flexoelectric coefficient for MoSi$_2$P$_4$, which is 0.041 nC/m. Considering CrSi$_2$P$_4$ material, which shows a flexoelectric coefficient of 0.033 nC/m and it is a little smaller as compared to MoSi$_2$P$_4$. Here also, we observed that Si atoms are producing similar dipole moments as in the case of MoSi$_2$N$_4$. The dipole moments from P atoms left and right to the Cr atom is cancelling out. Whereas, the top and bottom P atoms produce a difference of 0.0182 eÅ, which is about 0.86 times lower than that produced in MoSi$_2$P$_4$. This ratio is nearly identical to the flexoelectric coefficient ratio between CrSi$_2$P$_4$ and MoSi$_2$P$_4$. For the WSi$_2$P$_4$, the flexoelectric coefficient is only 0.017 nC/m. In addition to the similar observations made from MoSi$_2$P$_4$ and CrSi$_2$P$_4$, W core atom acquires higher $p_z$ compared Mo and Cr. Nonetheless, its direction is opposite to the resultant dipole moment from other atoms. This makes the total dipole moment of these system lower compared to others.

Among all studied MA$_2$Z$_4$ monolayers, WGe$_2$N$_4$ shows 0.047 nC/m flexo coefficient (find Table 3), which is by around 1.5 times higher than MoS$_2$ [76]. Here also W acquires a higher dipole moment $p_z$ due to the lower electron affinity. This dipole moment is in the direction of Ge dipole moment, similar to the case of MoSi$_2$N$_4$ shown in Fig. 8d. As a result, the total dipole moment variation is higher compared to other Ge based monolayers shown in Fig. 7d. An interesting point to note is, nanosheets terminated with N atoms exhibit high piezoelectric properties. The current asymmetry in upper and lower portions of the M core atom in the unitcell is not sufficient to further enhance the flexoelectric coefficient. But it appears that Janus type of MA$_2$Z$_4$ (like upper and bottom parts contain N, Si and P, Ge, respectively) may substantially enhance the flexo coefficient over Janus TMDs, which requires future examinations. When comparing the piezo coefficient of MA$_2$Z$_4$, they show superior piezoelectric coefficient over existing materials like MoS$_2$ [76,86,87], XTeI [88] and Janus TMDs [89,90]. A recent DFT study predicted that SnOSe Janus structure has the highest piezo coefficient of 1.12 nC/m [91] among the known 2D materials. Notably, our result for the piezo coefficient of MoSi$_2$N$_4$ is already by around 2 times higher than that of the SnOSe. More importantly, CrSi$_2$N$_4$ and WSi$_2$N$_4$ nanosheets even outperform MoSi$_2$N$_4$ (find Table 3). It can thus be concluded that MSi$_2$N$_4$ nanosheets record the highest piezoelectric coefficient among the all 2D materials. Nonetheless they show moderate flexoelectricity which can be very possibly enhanced by considering the Janus lattices.

**Table 3**, Total polarizability estimated by DFT ($\alpha_{DFT}$) and the present CD model ($\alpha_{CAL}$). Atomic polarizability ($R$) in 2H-*MA$_2$Z$_4$*, $d_{yyy}$ and $\mu_{yzyz}$ are the tensile piezoelectric coefficient and out-of-plane bending flexoelectric coefficient, respectively.

| Lattice | $\alpha_{DFT}$ (Å$^3$) | $\alpha_{CAL}$ (Å$^3$) | $R_M$ (Å) | $R_A$ (Å) | $R_Z$ (Å) | $d_{yyy}$ (nC/m) | $\mu_{yzyz}$ (nC/m) |
|---|---|---|---|---|---|---|---|
| MoSi$_2$N$_4$ | 36.372 | 35.968 | 1.006 | 1.386 | 0.505 | 2.293 | 0.007 |
| MoSi$_2$P$_4$ | 80.274 | 76.519 | 1.255 | 1.440 | 1.022 | 0.890 | 0.041 |
| MoGe$_2$N$_4$ | 41.023 | 40.913 | 1.223 | 1.443 | 0.520 | 1.775 | 0.004 |
| MoGe$_2$P$_4$ | 60.166 | 56.486 | 1.201 | 1.428 | 0.937 | 0.268 | 0.021 |
| CrSi$_2$N$_4$ | 35.452 | 35.063 | 1.097 | 1.409 | 0.650 | 3.687 | 0.009 |
| CrSi$_2$P$_4$ | 77.091 | 77.111 | 1.242 | 1.443 | 1.023 | 0.824 | 0.033 |
| CrGe$_2$N$_4$ | 33.282 | 32.727 | 1.224 | 1.470 | 0.693 | 3.163 | 0.004 |
| CrGe$_2$P$_4$ | 81.123 | 81.809 | 1.289 | 1.565 | 0.770 | 0.191 | 0.007 |
| WSi$_2$N$_4$ | 39.519 | 37.965 | 1.171 | 1.450 | 0.658 | 3.760 | 0.001 |
| WSi$_2$P$_4$ | 94.226 | 88.079 | 1.316 | 1.746 | 0.668 | 0.367 | 0.017 |
| WGe$_2$N$_4$ | 45.001 | 42.098 | 1.114 | 1.402 | 0.716 | 2.077 | 0.047 |
| WGe$_2$P$_4$ | 91.194 | 86.745 | 1.282 | 1.763 | 0.750 | 0.302 | 0.008 |

## 4. Concluding remarks

Motivated by the latest experimental advance in the fabrication of centimeter-scale high-quality single-layer MoSi$_2$N$_4$, in this work we conduct extensive first-principles simulations to explore the mechanical properties, lattice thermal conductivity, piezoelectric response and photocatalytic and electronic features of *MA$_2$Z$_4$* (M= Cr, Mo, W; A= Si, Ge; Z=N, P) monolayers. We show that depending on the atomic configuration and compositions, the studied nanomembranes show diverse electronic features from antiferromagnetic metal to half metal and to semiconductors with band gaps ranging from 0.31 to 2.57 eV. Interestingly, MoSi$_2$N$_4$, WGe$_2$N$_4$ and WSi$_2$N$_4$ nanosheets are found to exhibit appropriate band edge positions, excellent carrier mobilities and decent absorption of visible light and thus can be considered as promising candidates for the photocatalytic water splitting. The obtained results confirm remarkably high mechanical properties of studied nanosheets, in particular for the cases of MSi$_2$N$_4$ lattices. Notably, our results suggest outstanding thermal conductivities for the MSi$_2$N$_4$ nanosheets, reaching to around 440 and 500 W/mK for the MoSi$_2$N$_4$ and WSi$_2$N$_4$ monolayers, respectively, appealing for the thermal management systems. For the first time we show that machine learning interatomic potentials trained using 2000 time-step long AIMD trajectories over small 4×3×1 supercells enable the examination of flexoelectric and piezoelectric properties of complex structures. As the most exciting finding, WSi$_2$N$_4$, CrSi$_2$N$_4$ and MoSi$_2$N$_4$ are found to, respectively, record the highest piezoelectric coefficients, outperforming all other-known 2D materials. Our results clearly reveal the outstanding properties of MSi$_2$N$_4$ compositions. Therefore, in order to find the structure with the maximal mechanical, thermal conduction and piezoelectric properties, MSi$_2$N$_4$ (M=Mo, W, V, Nb, Ta, Ti, Zr, Hf or Cr) nanosheets should be further examined. Extensive results by this study highlight the exceptional physical properties of MA$_2$Z$_4$ nanomembranes, highly promising for the design of strong and robust nanoelectronics, optoelectronics, thermal management and energy conversion nanosystems. It is evident that this novel class of 2D materials not only outperforms the transition metal dichalcogenides group but can also compete with graphene in many applications.

## Acknowledgment

B.M. and X.Z. appreciate the funding by the Deutsche Forschungsgemeinschaft (DFG, German Research Foundation) under Germany's Excellence Strategy within the Cluster of Excellence PhoenixD (EXC 2122, Project ID 390833453). B.J. and X.Z. gratefully acknowledge the sponsorship from the ERC Starting Grant COTOFLEXI (No. 802205). Authors also acknowledge the support of the cluster system team at the Leibniz Universität of Hannover. B. M and T. R. are greatly thankful to the VEGAS cluster at Bauhaus University of Weimar for providing the computational resources. A.V.S. is supported by RFBR grant number 20-53-12012. F.S. thanks the Persian Gulf University Research Council for support of this study.

## Appendix A. Supplementary data

The geometry optimized structures and their corresponding PAW potentials in VASP native format are accessible via: http://dx.doi.org/10.17632/8v6n9mkbs5.1. The supplementary data to this article are provided after the references section.

Bohayra Mortazavi*[a,b], Brahmanandam Javvaji[a,#], Fazel Shojaei[c,#], Timon Rabczuk[d], Alexander V. Shapeev[e] and Xiaoying Zhuang[a,b,d]

[a]Chair of Computational Science and Simulation Technology, Institute of Photonics, Department of Mathematics and Physics, Leibniz Universität Hannover, Appelstraße 11,30157 Hannover, Germany.
[b]Cluster of Excellence PhoenixD (Photonics, Optics, and Engineering–Innovation Across Disciplines), Gottfried Wilhelm Leibniz Universität Hannover, Hannover, Germany.
[c]Department of Chemistry, College of Sciences, Persian Gulf University, Boushehr 75168, Iran.
[d]College of Civil Engineering, Department of Geotechnical Engineering, Tongji University, 1239 Siping Road Shanghai, China.
[e]Skolkovo Institute of Science and Technology, Skolkovo Innovation Center, Nobel St. 3, Moscow 143026, Russia.

*E-mail:  bohayra.mortazavi@gmail.com
#These authors contributed equally

Geometry optimized structures in VASP POSCAR and corresponding POTCARs of 2H- and 1T-MA$_2$Z$_4$ monolayers are accessible via: http://dx.doi.org/10.17632/8v6n9mkbs5.1 .

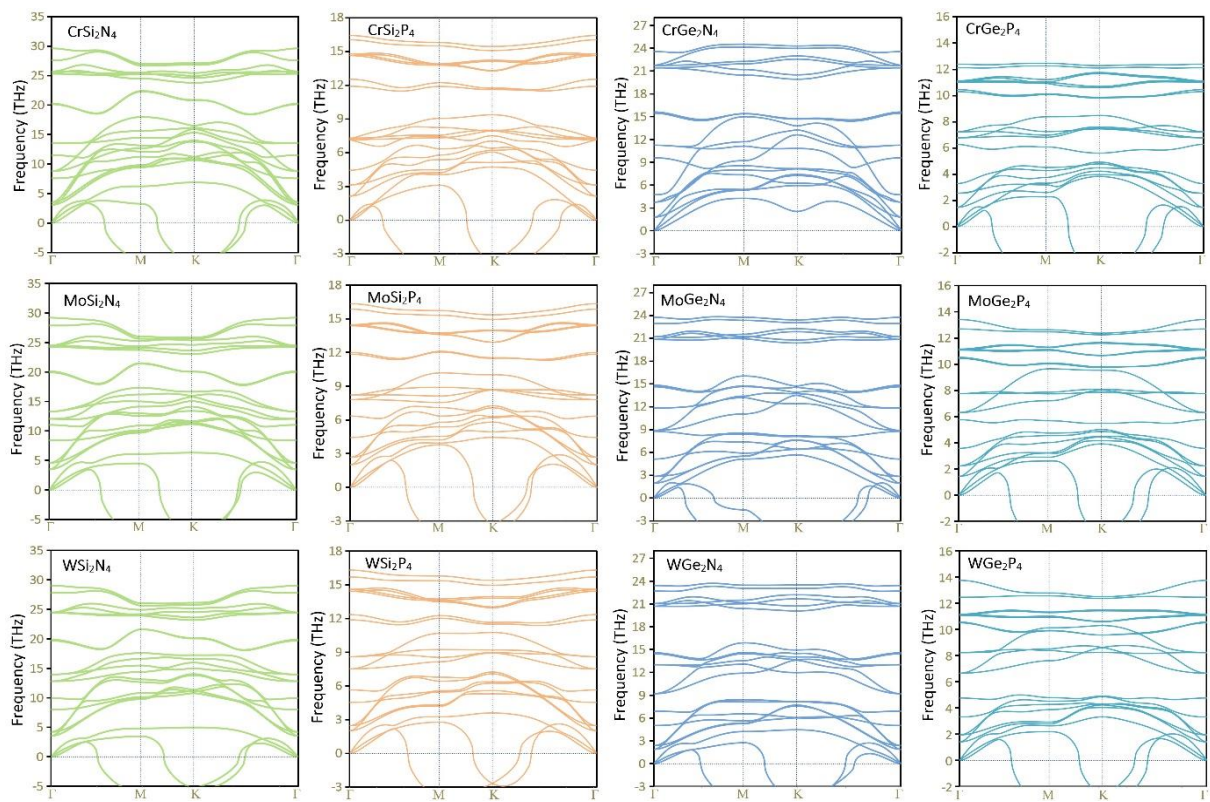

**Fig. S1**, Phonon dispersion relation of 1T-*MA$_2$Z$_4$* monolayer.

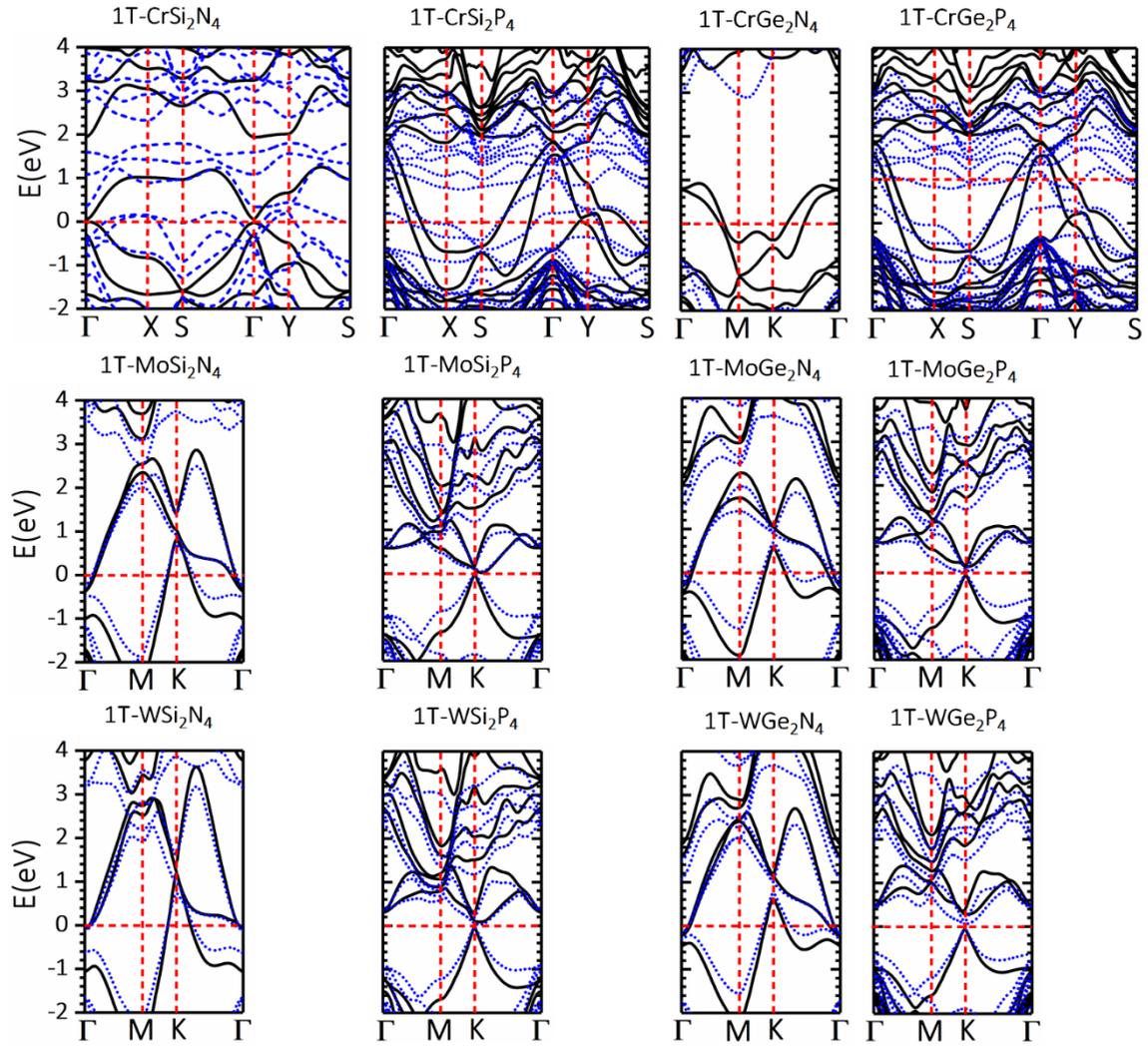

Fig. S2, HSE06 and PBE band structures for 1T-MA$_2$Z$_4$ monolayers in their magnetic ground states. Black solid and blue dotted lines represent HSE06 and PBE band structures, respectively. The Fermi level is set to 0 eV.

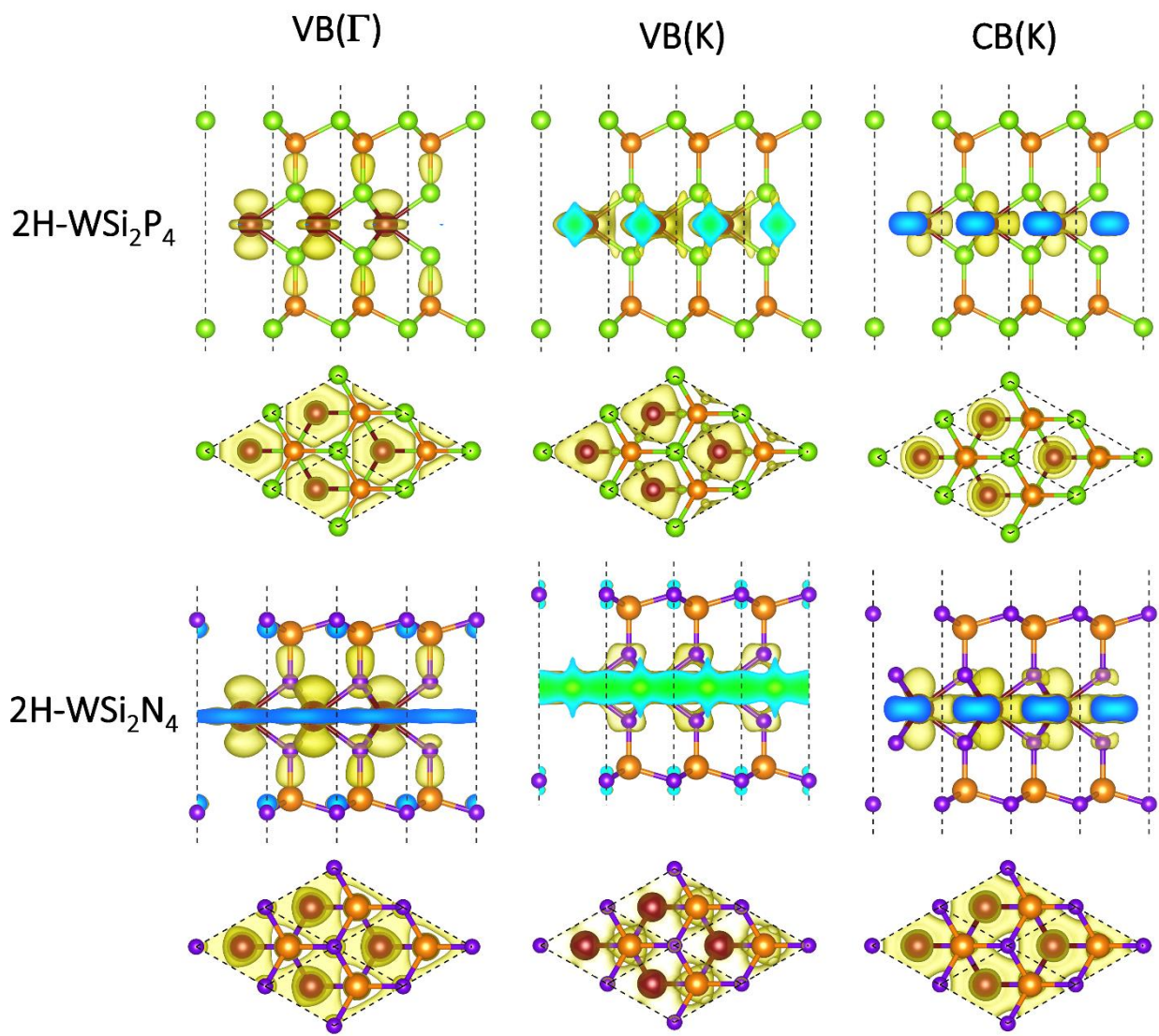

Fig. S3, Charge density distribution of two selected 2H-MA$_2$Z$_4$ monolayers at VB(Γ), VB(K) and CB(K). For comparison purposes isosurface value is set to 0.01 e/Å$^3$

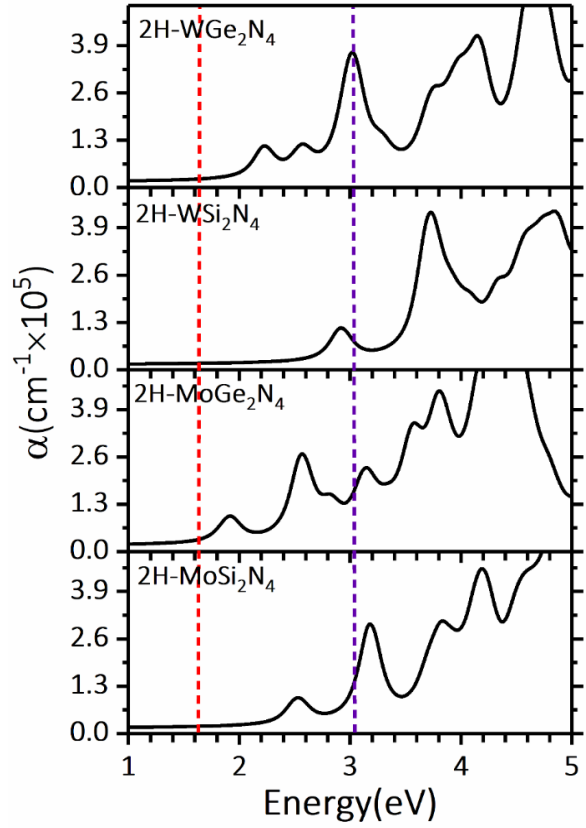

**Fig. S4,** Optical absorption coefficient obtained using HSE06 functional for 2H-WGe$_2$N$_4$, WSi$_2$N$_4$, MoGe$_2$N$_4$ and MoSi$_2$N$_4$ monolayers. The energy range of visible light spectrum is also shown by vertical dashed lines.

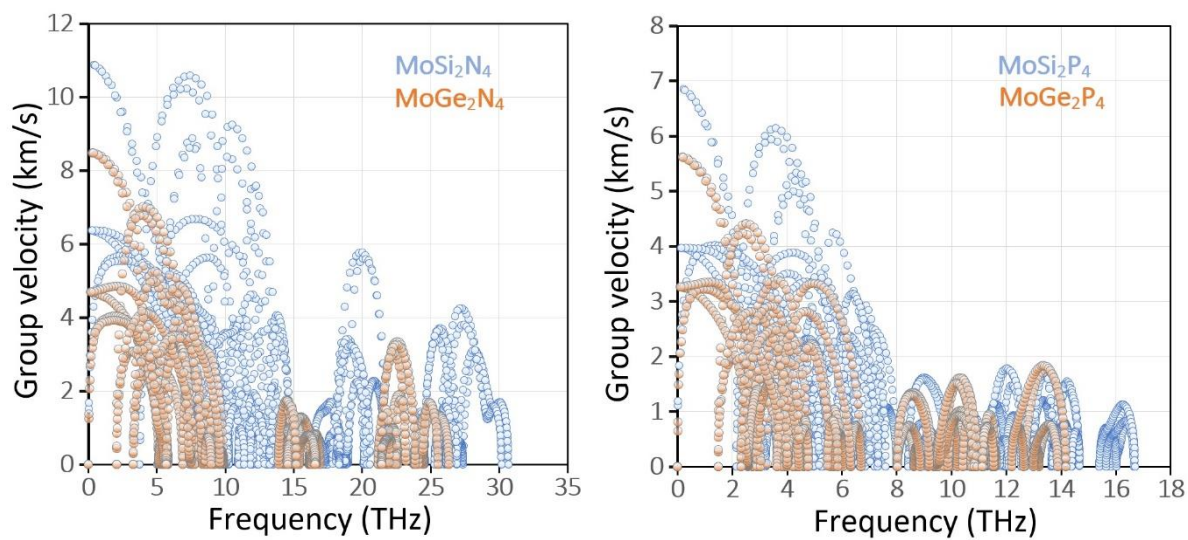

**Fig. S5**, Phonon group velocity of 2H-MoA$_2$Z$_4$ monolayers predicted using the PHONOPY with MTP.

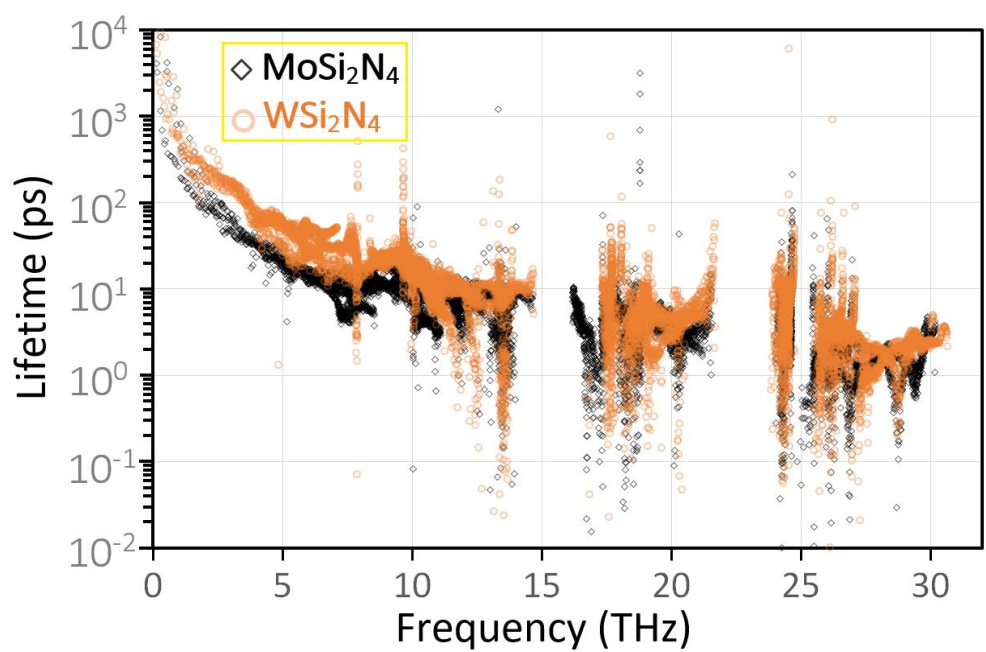

**Fig. S6**, Phonon lifetime of 2H-WSi$_2$N$_4$ and MoSi$_2$N$_4$ monolayers as a function of frequency acquired using ShengBTE with MTP.

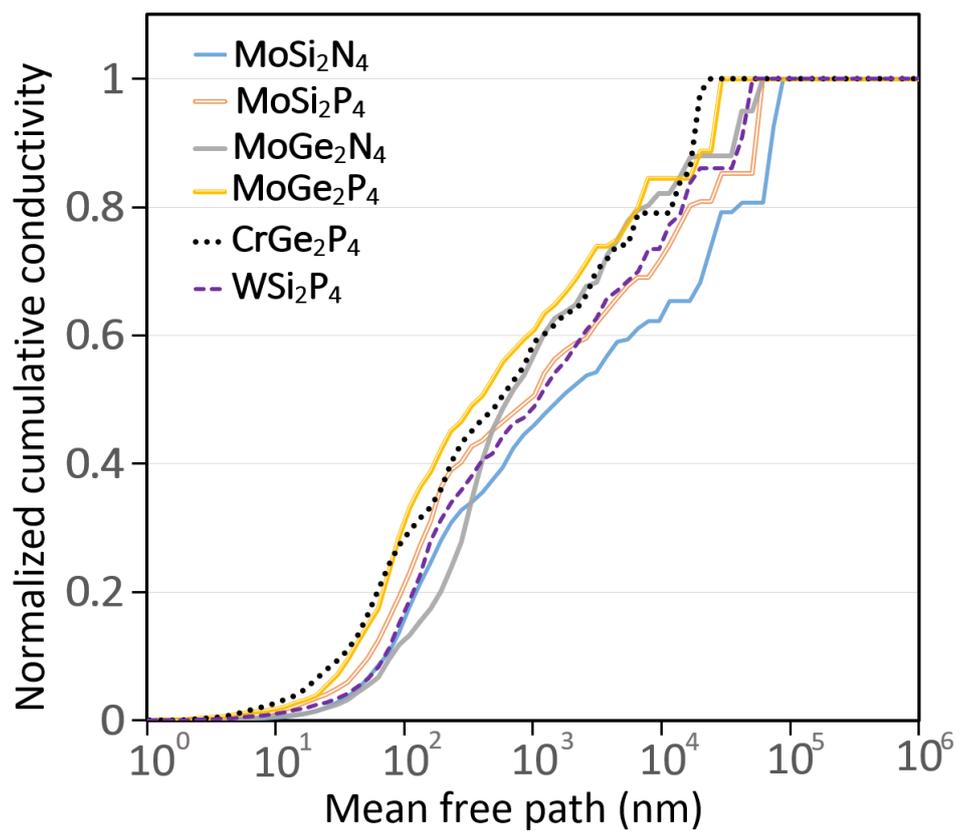

Fig. S7, Cumulative lattice thermal conductivity as a function of phonons mean free path.